\documentclass[a4paper,11pt]{article}
\usepackage{graphicx}
\usepackage{amsmath}
\usepackage{amssymb}
\usepackage{amsthm}
\usepackage{mathtools}
\usepackage{color}
\usepackage{url}
\usepackage{enumerate}
\usepackage{times}
\usepackage{a4wide}
\usepackage{pgf,tikz}
\usetikzlibrary{arrows}
\usetikzlibrary{patterns}
\usepackage{eufrak}

\newcommand{\nat}{\mathbb{N}}
\newcommand{\eoe}{} 
\newcommand{\eio}{\mathsf{eio}}
\newcommand{\model}{\mathsf{M}}
\newcommand{\ceiling}[1]{\lceil{#1}\rceil}
\newcommand{\floor}[1]{\lfloor{#1}\rfloor}
\newcommand{\prefix}{\mathit{pre}}
\newcommand{\nz}{\mathit{nz}}
\newcommand{\fz}{\mathit{fz}}
\newcommand{\Sys}{\mathfrak{D}}

\newtheorem{definition}{Definition}
\newtheorem{example}{Example}
\newtheorem{lemma}{Lemma}
\newtheorem{theorem}{Theorem}
\newtheorem{algorithm}{Algorithm}

\begin{document}

\title{A Theory of Black-Box Tests}
\author{Mohammad Torabi Dashti\footnote{
    Department of Computer Science.
    ETH Zurich.  Z\"urich, Switzerland. {\textit{mtorabi@inf.ethz.ch}}} \and David Basin\footnote{Department of Computer Science.
    ETH Zurich.  Z\"urich, Switzerland. {\textit{basin@inf.ethz.ch}}}
}
\date{}

\maketitle

\begin{abstract}
The purpose of testing a system with respect to a requirement is to
refute the hypothesis that the system satisfies the requirement.  We
build a theory of tests and refutation based on the elementary notions
of satisfaction and refinement. We use this theory to characterize the
requirements that can be refuted through black-box testing and,
dually, verified through such tests.  We consider refutation in finite
time and obtain the finite falsifiability of hyper-safety temporal
requirements as a special case.  We extend our theory with
computational constraints and separate refutation from enforcement in
the context of temporal hyper-properties. Overall, our theory provides
a basis to analyze the scope and reach of black-box tests and to
bridge results from diverse areas including testing, verification, and
enforcement.
\medskip

\end{abstract}

\section{Introduction}
\label{sec:intro}
\paragraph{Problem.}
In black-box testing, the internal structure of the system under test,
including its hardware and the algorithms and data structures
implemented, are unknown to the tester.  The need for black-box
testing arises when testers have no access to, or auxiliary
information about, the system under test, other than what they
can observe by interacting with the system over its interface, e.g.,
by providing the system with inputs and observing
its outputs.

Despite the simplicity of the black-box setting and the manifest
importance of testing in general, the theory of black-box testing is
under-developed and a solid understanding of its strength and
limitations is lacking.  For instance, it is commonly agreed upon that
the purpose of testing a system with respect to a requirement is to
refute the hypothesis that the system satisfies the
requirement~\cite{dijkstra70,myers11}. Yet existing testing theory is
inadequate for answering basic questions in the black-box setting such
as: which class of requirements are refutable, given a class of tests?
Or, which class of tests, if any, can refute a class of requirements?

We develop a theory of black-box testing that explicates what can be
determined about systems by observing their behavior.  Our theory
fully characterizes the class of refutable and verifiable
requirements.  This means it precisely specifies for which system
requirements the violation (respectively satisfaction) can, or cannot,
be demonstrated through black-box tests.
Establishing the limits of testing this way is analogous to
establishing elementary results in complexity theory that delimit the
boundaries of effective computation.  Moreover, our theory helps
testers understand the consequences of implicit assumptions they may
be making when carrying out tests, for example, that systems are
deterministic.  Our theory also provides a foundation for bridging
results in testing with related disciplines.

\paragraph{Approach.}
We start with an abstract model of systems and requirements
(\S\ref{sec:refut}) and introduce two types of requirements: obligations
and prohibitions (\S\ref{sec:reqtype}).  A requirement is an obligation
if it obliges the systems to exhibit certain (desired) behaviors, and it
is a prohibition if it prohibits the systems from exhibiting (undesired)
behaviors.  Here, a behavior could be, for example, sets of input/output
pairs, or sets of traces.  Functional requirements are typically
obligations, and security requirements are, by and large,
prohibitions. We show that these two requirement types admit a
straightforward order-theoretic characterization.  Namely, given a
refinement (or abstraction) partial-order on a set of systems, the
satisfaction of an obligation is abstraction-closed, and for a
prohibition it is refinement-closed.

We turn next to black-box tests (\S\ref{sec:bbt}).  Given a black-box
system, the tester can observe its input and output, but cannot
observe \emph{how} the latter is produced from the former.  The tester
can therefore analyze such a system only by interacting with it over
its interface, and not by, for example, analyzing its software.  In
black-box testing, sometimes called ``testing by
sampling''~\cite{dijkstra70}, testing amounts to inspecting a sample
of system behaviors.  The sample obtained through tests can be seen as
a refinement of the system under test, a notion we make precise in
subsequent sections.  All a tester learns by sampling is that the
system exhibits certain behaviors. From this, the tester cannot infer
that the system does not exhibit other behaviors as well. Such a
conclusion could only be justified through the sample's
exhaustiveness, which black-box testing alone cannot establish.  A
requirement is therefore refutable through tests if, for any system
that violates the requirement, the hypothesis that the system
satisfies the requirement can be refuted by inspecting a refinement of
the system.

It follows that a requirement whose violation is contingent upon
demonstrating the absence of behaviors cannot be refuted through
black-box testing.  Based on this, we prove that any refutable
requirement is a prohibition, and all non-trivial obligations are
irrefutable (\S\ref{sec:refreq}).  We then define the notion of
verification dual to refutation, and show that any verifiable
requirement is an obligation and that non-trivial prohibitions cannot
be verified through tests (\S\ref{sec:verif}).

The theory sketched above is aimed to delimit the scope and reach of
black-box testing in general.  However, it does not account for two
central limitations of black-box tests in practice: testing must proceed
in a finite amount of time, and test oracles must be computable. We
specialize our theory to accommodate these limitations. Namely,
we introduce the notions of \emph{finite refutability} (and dually
\emph{finite verifiability}), and characterize the class of finitely
refutability (verifiable) requirements (\S\ref{sec:temporal}). Our main
result here relates finitely refutable temporal requirements to safety
properties and hyper-safety hyper-properties. We further specialize
the theory by considering the case when test oracles are constrained to
be computable (\S\ref{sec:alg}).  We use this specialization to separate
properties that are refutable from those that are enforceable by runtime
monitoring.

While \S\ref{sec:temporal} and \S\ref{sec:alg} pertain to
  specializations of our black-box testing theory, in \S\ref{sec:aux} we consider a
generalization: testing in the grey-box setting where testers may have
partial information about the system under test.  It is not surprising
that access to auxiliary information can enlarge the set of refutable
requirements. Our main result here is to show that refutation with the
help of auxiliary information can be reduced to the task of refuting a
prohibition. This result further illustrates the tight connection between
prohibitions and refutability mentioned above.  We also use this generalization
to explicate the assumptions that are implicit in
several well-known testing techniques (\S\ref{sec:practice}). 

Overall, we present a basic theory for reasoning about the strength
and limitations of black-box testing. Our theory is abstract and
has minimal formal machinery, which makes it easy to extend.  We
present specializations and generalizations that account for refutation in
finite time, refutation under computational constraints, and
refutation aided by auxiliary information.  We use these to prove new
results and to obtain known results as special cases, as
explained in the following.

\paragraph{Contributions.}  
Our first contribution is the theory sketched above.  We use it to
fully characterize the requirements that can be refuted and those that
can be verified through black-box tests. 
Our proofs are short;  they often amount to simply
  unrolling definitions. This suggests that our theory is at the
  right level of abstraction for reasoning about black-box tests, a
  claim which is further supported by observing that the theory lends
  itself to direct, straightforward extensions, as discussed
  above.

Our second contribution is to show how our theory can be used to
derive both known and new results in a straightforward way.  In
particular, we present different applications of our theory of finite
refutability (\S\ref{sec:temporal}).  For instance, we demonstrate
that the finite falsifiability of hyper-safety temporal
requirements established by Clarkson and Schneider~\cite{DBLP:journals/jcs/ClarksonS10}
can be derived as a special case in our theory.  As
another example, we use our characterization to separate refutability
from enforceability: we show that any enforceable temporal requirement
is refutable, but refutable requirements need not be enforceable.  
This separation hinges upon analyzing the
computational constraints of refutation (and enforcement) via a notion
of algorithmically refutable requirements (\S\ref{sec:alg}). Moreover,
the abstract nature of our theory allows us to establish connections
between algorithmic refutability, topology, and recursion theory.

Our third contribution is to use our characterization of refutability
through black-box tests to augment and shed light on the folkloric
understanding of testing that exists in the community.  As an example,
consider Dijkstra's statement that ``program testing can be used to
show the presence of bugs, but never to show their absence'', which is
widely quoted in software testing community.  We make precise a
stronger version of this statement, and show that its proof is
independent of the cardinality of the input domain, i.e., the number
of test cases one must consider.  Moreover, we use our
characterization to rectify the folklore surrounding Dijkstra's
statement, for example, that testing can never be used to establish
that a system satisfies a requirement.  As a second example, we
highlight the fundamental role that determinacy assumptions play in
making sense of day-to-day black-box functional tests
(\S\ref{sec:aux}).  In particular, we examine three prominent testing
techniques, namely functional testing, model-based testing, and fuzz
testing, in light of our theory, and explicate their implicit
assumptions (\S\ref{sec:practice}).
We discuss other related work in~\S\ref{sec:relwork} and conclude by
discussing the limitations of our theory in~\S\ref{sec:conc}.

Parts of the work described here were published
in~\cite{DBLP:conf/atva/DashtiB17}. The current article extends this
previous work with additional technical details, examples, and
explanations, pertaining to the notions of refutability,
verifiability, and black-box testing. Moreover, the notion of
refutability under auxiliary assumptions (\S\ref{sec:aux}) as well as
the systematic review of testing practice (\S\ref{sec:practice}) are
entirely new.

\section{Systems and Requirements}
\label{sec:refut}
We give a simple abstract model of systems and requirements, the main
ingredients of our theory.

A system is an entity that is capable of exhibiting
observable behaviors.  Operating systems, digital circuits and vending
machines are all examples of systems.  We keep the notion of an
observable behavior unspecified for now and instead work with systems
as a set of objects with an associated partial order.  Namely,
let~$\Sys$ denote the nonempty set of all systems under consideration,
which is our domain of discourse.  We assume that~$(\Sys,\preceq)$ is
a partially-ordered set (poset), where~$\preceq$ denotes a refinement
relation:~$S_1\preceq S_2$ means that~$S_2$ exhibits all the behaviors
of~$S_1$. In this case, we say system~$S_1$ {\bf refines}
system~$S_2$, or system~$S_2$ {\bf abstracts} system~$S_1$.

There exists a large body of research on refinement and abstraction;
see for instance~\cite{DBLP:conf/lics/AbadiL88,morgan,glabbeek90}.
Examples of refinement relations include trace containment and various
algebraic simulation relations.  In the interest of generality, we do
not bind~$\preceq$ to any particular relation.  We write~$\ceiling{S}$
and~$\floor{S}$ respectively for the set of systems that abstract a
system~$S$ and those that refine it: $\ceiling{S}=\{S'\in
\Sys\mid S\preceq S'\}$ and~$\floor{S}=\{S'\in \Sys\mid
S'\preceq S\}$.  We assume that the poset~$(\Sys,\preceq)$ is
bounded: it has a greatest element~$\top$ and a least element~$\bot$.
The ``chaos'' system~$\top$ (sometimes called the ``weakest''
system~\cite{unifying}), abstracts every system, and the ``empty''
system~$\bot$ refines every system in~$\Sys$.  In short, our
{\bf system model} is a four-tuple~$(\Sys,\preceq,\bot,\top)$.

We remark that $\top$ and $\bot$ are fictitious entities in the sense
that there is no need to construct them. We will use $\top$ 
to reason about the testers' epistemic limitations.  In contrast,
$\bot$ is, strictly speaking, not necessary for our theory's
development. We introduce it for the sake of symmetry and
as a shorthand for ``empty'' systems.

We extensionally define a {\bf requirement} to be a set of systems. A
system {\bf satisfies} a requirement~$R$ if it belongs to~$R$. For
example, the requirement stipulating that systems are deterministic
consists of all the deterministic systems in $\Sys$. For now, we need
not expound on the satisfaction relation between systems and
requirements; we will give examples later.  We write~$\chi_R$ for a
requirement~$R$'s characteristic function, which maps~$\Sys$ to
$\{0,1\}$.  A requirement~$R$ is {\bf trivial} if all or none of the
systems in $\Sys$ satisfy it, i.e.~$\chi_R$ is a constant function iff
$R$ is trivial.

It is immediate that~$(\mathcal{R},\subseteq)$ is a complete lattice,
where~$\mathcal{R}$ is the set of all requirements and~$\subseteq$ is the standard set
inclusion relation. We define the {\bf
  conjunction} of two requirements~$R_1$ and~$R_2$, denoted~$R_1\wedge
R_2$, as their meet, and their {\bf disjunction}, denoted $R_1 \vee
R_2$, as their join. For a nonempty set~$R$ of systems, we
write~$\ceiling{R}=\bigcup_{S\in R}\ceiling{S}$
and~$\floor{R}=\bigcup_{S\in R}\floor{S}$.  A set~$R$ is {\bf
  abstraction-closed} if~$R=\ceiling{R}$, and {\bf refinement-closed}
if~$R=\floor{R}$. Such a set is called an upper set, and respectively,
a lower set in order theory.

\section{Requirement Types}
\label{sec:reqtype}
We define obligations and prohibitions, and prove a lemma that
separates these requirement types (\S\ref{sec:oblpro}). Afterward, we
characterize the requirements that can be expressed as the conjunction
of an obligation and a prohibition (\S\ref{sec:semi}). Finally, we
present an intuitive interpretation of obligations and prohibitions
as, respectively, lower-bounds and upper-bounds on system behaviors
(\S\ref{sec:interp}).

\subsection{Obligations and Prohibitions}
\label{sec:oblpro}
A requirement is an obligation if it obliges the systems to exhibit
certain (desired) behaviors, such as intended
functionalities and features.  For example, a requirement for a
database system obliges it to provide the user with an option to
commit transactions.  This requirement cannot be violated
by adding behaviors to the system, for example by providing the user
the option to review transactions.  The satisfaction of an
obligation~$R$ is therefore abstraction-closed:~$\forall S,
S'\in\Sys.\ S\in R\ \wedge\ S\preceq S'\ \to\ S'\in R$.  

A requirement is a prohibition if it prohibits the systems from
exhibiting certain (undesired) behaviors.  For instance, consider the
requirement that prohibits a database system from committing malformed
transactions.  This requirement cannot be violated by
removing behaviors from the system, for example removing the option
for committing transactions altogether.  That is, the satisfaction of
a prohibition~$R$ is refinement-closed: $\forall
S,S'\in\Sys.\ S\in R\ \wedge\ S'\preceq S\ \to\ S'\in R$.  

Rewriting the previous two formulas gives us the following definition.

\begin{definition}
\label{def:obl-proh}
A requirement~$R$ is an {\bf obligation} if~$R=\ceiling{R}$, and~$R$
is a {\bf prohibition} if~$R=\floor{R}$.
\end{definition}

The following example illustrates the system model
 of~\S\ref{sec:refut},
obligations, and prohibitions. The example also introduces the {\bf
  extensional input-output} system model $\eio{}$, which we use
throughout the paper.  This model highlights two features of interactive
systems that (1) distinguish between inputs and outputs,
and (2) react to any input, either by producing an output
(including undesired ones, such as throwing an exception or crashing)
or by diverging, i.e.\ not terminating.

\begin{example}
\label{eio-def-ex} 
Consider the system
model~$(2^{\nat\times\nat},\subseteq,\emptyset,\nat\times\nat)$, where
a system is extensionally defined as a subset of~$\nat\times\nat$,
with~$\nat$ being the set of natural numbers, and the refinement
relation is the standard subset relation. For an input~$i\in\nat$, a
system~$S$ produces an output~$o$, non-deterministically chosen from
the set~$\{n\in\nat\mid (i,n)\in S\}$, and it does not produce any
outputs when~$\{n\in\nat\mid (i,n)\in S\}$ is empty.  We call this
system model~$\eio{}$.  Note that, due to its extensional definition,
this model makes no distinctions between two systems that define the
same subset of $\nat\times\nat$ but are otherwise different, e.g., one
of them runs faster than the other.

The requirement~$P$ stipulating that systems are deterministic is a
prohibition: if~$S$ is deterministic, meaning~$\forall
i\in\nat.\ |\{n\in\nat \mid (i,n)\in S\}|\le 1$, then so is any
refinement, i.e.\ subset, of~$S$. In particular, the empty system
satisfies the definition of determinacy.

The requirement~$O$ stipulating that systems define total relations is
an obligation: if~$S$ is total, meaning~$\forall
i\in\nat.\ |\{n\in\nat \mid (i,n)\in S\}|> 0$, then so is any
abstraction, i.e.\ superset, of~$S$. In particular, $\top$ is total.

The requirement~$R$, stating that systems extensionally define total
functions, is clearly neither a prohibition nor an obligation:
from~$\forall i\in\nat.\ |\{n\in\nat \mid (i,n)\in S\}|=1$ we cannot
conclude that an arbitrary subset or superset of~$S$ defines a total
function. Note that~$R=O\wedge P$.~\eoe
\end{example}

Two remarks are due here. First, Definition~\ref{def:obl-proh}
qualifies the relationship between a requirement's
satisfaction and the notion of refinement. Analogous formulations are
found, for example, in logic. A satisfiable sentence remains
satisfiable after enlarging the set of models, whereas a valid
sentence remains valid after reducing the set of models.
In this sense,  obligations resemble satisfiability, and prohibitions resemble
validity.

Second, syntactically reformulating a requirement's description does
not affect its type.  For example, the prohibition stating that
\emph{systems may not produce two (or more) different outputs for any
  input} can be syntactically reformulated as \emph{systems may
  produce at most one output for each input} without affecting its
type. The latter formulation permits, and the former forbids, certain
behaviors. As a second example, the requirement $F$ 
that forbids ``doing nothing'' is abstraction-closed, simply because all systems
except $\bot$ satisfy $F$. That is, $F$ is an obligation, in spite of
the term ``forbid'' appearing in its statement. In short, the
syntactic disguise of a requirement plays no role in determining its
type. 

We now separate obligations and prohibitions by showing that a
nontrivial requirement cannot belong to both these types.
A requirement~$R$ is an obligation iff~$\chi_R$ is monotonically
increasing in~$\preceq$, that is,~$S\preceq S'\to \chi_R (S)\le
\chi_R(S')$.  Similarly,~$R$ is a
  prohibition iff~$\chi_R$ is monotonically decreasing, that
  is,~$S\preceq S'\to \chi_R(S')\le \chi_R(S)$.  Therefore, any
  requirement that is both an obligation and a prohibition must have a
  constant characteristic function and hence is trivial. The following
  lemma is now immediate. 
(See Appendix~\ref{app-proof} for all proofs.) 

\begin{lemma}
\label{lem:distinct}
If a requirement~$R$ is both an obligation and a prohibition, then~$R$
is trivial.
\end{lemma}

This lemma implies that a prohibition cannot be replaced with an
obligation and vice versa. For example, the prohibition \emph{smoking
  is forbidden} has no equivalent obligation, and the obligation
\emph{sacrifice a ram} has no equivalent prohibition.  The lemma does
not however imply that obligations and prohibitions exhaust the set of
requirements. A {\bf non-monotone} requirement, i.e.\ one
whose characteristic function is neither monotonically increasing nor
monotonically decreasing, is neither an obligation nor a
prohibition. For instance, the requirement~$R=O\wedge P$, defined in
Example~\ref{eio-def-ex}, is neither an obligation nor a prohibition,
as it is not monotone. 

Many practically-relevant requirements turn out to be the conjunction
of an obligation and a prohibition, similarly to~$R$ above. We
generalize this to the notion of semi-monotonicity.

\subsection{Semi-Monotonicity}
\label{sec:semi}
A requirement is {\bf semi-monotone} if it is the conjunction of two
(or more) monotone requirements.  In~\S\ref{sec:refreq} we show that,
when it comes to refutability, a semi-monotone requirement behaves
like a monotone one, but only for \emph{some} systems.  This motivates
studying semi-monotonicity.

Semi-monotonicity is strictly weaker than monotonicity. Consequently,
obligations and prohibitions are (trivially) semi-monotone, but a
semi-monotone requirement need not belong to either of these types.  The
following lemma states that obligations and prohibitions, closed under
conjunction, are necessary and sufficient for expressing all
semi-monotone requirements. Note that, due to the idempotence
of~$\ceiling{\cdot}$ and~$\floor{\cdot}$, $\ceiling{R}$ is an
obligation and $\floor{R}$ a prohibition for any requirement~$R$.
\begin{lemma}
\label{lem:semi}
A requirement $R$ is semi-monotone iff $R = \ceiling{R} \wedge
\floor{R}$.
\end{lemma}

Although semi-monotonicity holds for many requirements, not all
requirements are semi-monotone, as the following example illustrates.

\begin{example}
\label{ex:zigzag}
Consider the~$\eio{}$ model and the requirement~$R$ that is satisfied
by a system~$S$ if for each $(i,o)\in S$ there exists some $(i', o)\in
S$, with~$i\neq i'$.  This requirement, which can be seen as a
simplified form of a~$k$-anonymity requirement~\cite{sweeney}, states
that by solely inspecting a system's outputs, an observer cannot
determine whether or not the input is some particular~$i\in\nat$.

Consider the ascending chain of systems $S_0\preceq S_1\preceq
S_2\preceq\cdots$, where $S_0=\{(0,0)\}$, and
$S_j=S_{j-1}\cup\{(j,o)\}$, where~$o=j/2$ if~$j$ is even,
and~$o=(j-1)/2$ otherwise.
That is,~$S_1=S_0\cup\{(1,0)\}$,~$S_2=S_1\cup\{(2,1)\}$, and so forth.
Note that~$S_j$ satisfies~$R$ iff~$j$ is odd, with~$j\in\nat$.  The
diagram below illustrates~$\chi_R$ with respect to the systems'
indices in the chain.

 
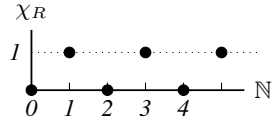
\begin{center}
\begin{tikzpicture}
\draw[line width=.24mm] (0,0) -- (0,0.8);
\node [above] at (0,.8) {\footnotesize{$\chi_R$}};
\draw[line width=.24mm] (0,0) -- (2.8,0);
\node [right] at (2.8,0) {\footnotesize{$\nat$}};
\draw[dotted] (0,.5) -- (3,0.5);
\node [left] at (0,0.5) {\footnotesize{1}};
\node [below] at (0,0) {\footnotesize{0}};
\draw (.5,0) -- (.5,.1);
\draw (1,0) -- (1,.1);
\draw (1.5,0) -- (1.5,.1);
\draw (2,0) -- (2,.1);
\draw (2.5,0) -- (2.5,.1);
\node [below] at (0.5,0) {\footnotesize{1}};
\node [below] at (1,0) {\footnotesize{2}};
\node [below] at (1.5,0) {\footnotesize{3}};
\node [below] at (2,0) {\footnotesize{4}};
\draw[fill=black] (0,0) circle (.07);
\draw[fill=black] (.5,.5) circle (.07);
\draw[fill=black] (1,0) circle (.07);
\draw[fill=black] (1.5,0.5) circle (.07);
\draw[fill=black] (2,0) circle (.07);
\draw[fill=black] (2.5,0.5) circle (.07);
\end{tikzpicture}
\end{center}

It is easy to check that on any chain~$S_0\preceq S_1 \preceq \cdots$,
a monotone requirement's characteristic function changes its value at
most once. A semi-monotone requirement's characteristic function
changes at most once from 0 to 1, and at most once from 1 to 0.  From
the zigzagging~$\chi_R$ of the above diagram, it is evident that~$R$
is not semi-monotone.

Alternatively, note that $\ceiling{R}\wedge \floor{R}=\Sys\neq
R$, and hence $R$ is not semi-monotone due to
Lemma~\ref{lem:semi}.~\eoe
\end{example}

We conclude this section by remarking that semi-monotonicity is
invariant under conjunction: $\bigwedge_{R\in \rho} R$ is
semi-monotone for any nonempty set~$\rho$ of semi-monotone
requirements (see the appendix for details).
This justifies the practice of piece-wise specification of
(semi-monotone) requirements, e.g.\ one for negative integers and one
for non-negative ones, and then combining them with conjunction.
Note however that semi-monotonicity is not an invariant under
disjunction because \emph{any} nonempty requirement~$R$ is the
disjunction of (infinitely many) semi-monotone requirements:
$R=\bigvee_{S\in R}\ceiling{S}\wedge \floor{S}$.

\subsection{Lower-Bound and Upper-Bound Interpretations}
\label{sec:interp}
Here we present an intuitive interpretation of obligations and
prohibition that we illustrate using the $\eio$ model.  In the
$\eio{}$ model, an obligation~$O$ is a set of systems that can be
characterized also by a set of desired behaviors.  Any violation
of~$O$ is therefore due to the behaviors that~$S$ lacks. Consequently,
$O$ can be seen as a lower-bound for the set of~$S$'s behaviors.
Similarly,~$S$ satisfies a prohibition~$P$ iff the set of behaviors
of~$S$ is contained in the set of behaviors~$P$ permits. Any violation
of~$P$ is therefore due to excessive behaviors of~$S$. In this
sense,~$P$ gives rise to an upper-bound for the set of~$S$'s
behaviors.

\begin{example}
\label{ex:ooi} 
Consider the $\eio{}$ system model, and the obligation~$O$ stating
that a (non-deterministic) system~$S$
must  produce $i+1$ as one of its possible outputs for every even input~$2i$.
Then, $S\in O\ \leftrightarrow\ F\subseteq S$, where
$F=\{(2i,i+1)\mid i\in \nat\}$.
Now, consider the prohibition~$P$ stating that a system may only
output~$i+1$ for an even input~$2i$, and for odd numbers the system
never outputs~0. Then, $S\in P\ \leftrightarrow\ S\subseteq G$, where
$G=F\cup\{(2i+1,o+1)\mid i,o\in\nat\}$.

We can now express the satisfaction relation between~$S$
and~$R=O\wedge P$ as $S\in R\ \leftrightarrow\ F\subseteq S\subseteq
G$.  A system violates~$R$ iff it lacks a behavior of~$F$, or it
exhibits a behavior outside~$G$.~\eoe
\end{example}

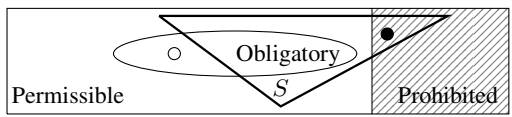
\begin{figure}[t]
\begin{center}
\begin{tikzpicture}
\draw[pattern=north east lines, pattern color=gray] (1.4,-1.2) -- (3.2,-1.2) -- (3.2, .2) -- (1.4, .2);
\draw[fill=white]  (-3.4,-1.2) -- (1.4,-1.2) -- (1.4,.2) -- (-3.4,.2) -- (-3.4,-1.2) ;
\draw (-.4,-.4) ellipse (1.6 and 0.3);
\draw [line width = .3mm] (-1.4,.1) -- (2.4,0.1) -- (0.2,-1.1) -- (-1.4,.1);
\draw (-1.2,-.4) circle (.08);
\draw [fill=black](1.6,-.14) circle (.08);
\node [above] at (2.4,-1.2) {\footnotesize Prohibited};
\node [above] at (-2.6,-1.2) {\footnotesize Permissible};
\node [above] at (0.3,-0.7) {\footnotesize Obligatory};
\node [above] at (0.2,-1.1) {\footnotesize{$S$}};
\end{tikzpicture}
\end{center}
\caption{The set of all behaviors is partitioned into the set of
  prohibited behaviors, represented by the hatched area, and the 
  set of permissible ones, represented by the white box. The set of
  obligatory behaviors, represented by the oval, is included in the
  set of permissible behaviors.  The triangle stands for a
  system~$S$'s behaviors.  The white circle represents a violation of
  the obligation denoted by the oval, and the black circle represents
  a violation of the prohibition depicted by the hatched area.}
\label{fig:pos-neg-ex}
\end{figure}

The diagram of Figure~\ref{fig:pos-neg-ex} illustrates the lower-bound
and upper-bound interpretations, where the oval is the lower-bound and
the white box is the upper-bound on system behaviors.
A similar figure is given
in~\cite{fuzzing-book}. 

Two remarks are due here. First, interpreting
prohibitions as a set of prohibited behaviors leads to a natural
definition of permissible behaviors.  Namely, the set of {\bf
  permissible behaviors} complements the set of prohibited ones,
cf.\ deontic logic~\cite{deontic}.  To avoid inconsistency, all
obligatory behaviors must be permissible, but not all permissible
behaviors need be obligatory.  Consequently, the set of permissible
behaviors for a system, delimited by the prohibitions, does not
necessarily coincide with its set of obligatory behaviors, as
illustrated in Figure~\ref{fig:pos-neg-ex}.

Second, a requirement $R$ that is not semi-monotone does not admit the
lower-bound and upper-bound interpretations: a system may violate $R$
even when it is bounded from below and from above by systems that
satisfy $R$. As Example~\ref{ex:zigzag} shows, $S_1\in R$, $S_3\in R$,
and $S_1\preceq S_2 \preceq S_3$ do not entail $S_2\in R$.  

\section{Black-Box Tests}
\label{sec:bbt}

Recall that a system is a black-box if we can observe its input and
output, but cannot observe how the latter is produced from the former.
In black-box testing, a tester can only interact with the system over
its interface.  We now characterize this in our system model.  We
start by defining a test setup, which enables us to distinguish system
behaviors from what a tester observes.

Let~$(\Sys,\preceq,\bot,\top)$ be a system model.  By sampling the
behaviors of a system~$S\in\Sys$, a tester makes an {\bf
  observation}.  We do not further
specify observations. We give examples shortly.  A {\bf test setup} is
a pair~$(T,\alpha)$, where~$T$ is a domain of observations
and~$\alpha: \Sys \to 2^T$ is an {\bf order-preserving} function,
i.e.,~$S\preceq S'\ \to\ \alpha(S)\subseteq \alpha(S')$.

Intuitively,
the set~$\alpha(S)$ consists of all the observations that can be made
by testing a system~$S$ in this test setup and a black-box test
simply amounts to making an observation from this set.
Since~$\alpha$ is
order-preserving, if~$t$ belongs to $\alpha(S)$ for some system~$S$,
then~$t\in\alpha(S')$ for any system~$S'$ that abstracts~$S$. This
reflects the nature of black-box testing where analyzing a system~$S$
``by sampling'' amounts to inspecting a sample of~$S$'s
behaviors~\cite{dijkstra70}. Therefore, if an observation can be made
on~$S$ by inspecting the behaviors $S$ exhibits, then the same
observation can also be made on any system~$S'$ that abstracts~$S$,
simply because~$S'$ exhibits all of~$S$'s behaviors. We illustrate
these notions with an example.

\begin{example}
\label{ex:cir}
Consider the $\eio{}$ system model and the test
setup~$\mathbf{T}_r=(\Sys,\floor{\cdot})$, where a tester may observe
an arbitrary refinement of the system under test.  Note that~$\floor{\cdot}$ is order-preserving
and hence~$\mathbf{T}_r$ is a test setup. The subscript $r$ indicates
\emph{reflexivity}: a system $S$ is itself a legitimate observation on
$S$ in the test setup $\mathbf{T}_r$.  In Section~\ref{sec:refutbbt},
we explain reflexive test setups in detail.  

Suppose a tester observes that a system~$S$ outputs 0 on input 0, and
1 on input 1. That is, the tester makes the
observation~$t=\{(0,0),(1,1)\}$ on~$S$, which is a refinement of $S$.
Clearly $\top$ could also
yield~$t$, simply because $t\in \floor{\top}$.~\eoe
\end{example}

We define the function~$\hat\alpha:T\to 2^\Sys$ to map 
an observation to the set of systems that can yield that
observation. Formally,~$\hat\alpha(t)=\{S\in \Sys\mid t\in
\alpha(S)\}$, for any~$t\in T$. In black-box testing, a tester knows
nothing about the behaviors of the system under test beyond what is
observed by interacting with it. Therefore, all the tester can
conclude from an observation~$t$ is that the system under test can be
\emph{any} system that could yield~$t$.  That is, solely based on an
observation~$t$, the tester cannot distinguish between the system
under test and any other system in~$\hat\alpha(t)$. We call
this the {\bf indistinguishability condition}.

The indistinguishability condition can be seen as providing an
epistemic basis for the standard structurally-oriented definition of
black-box testing given in the introduction.
In fact, the key observation enabling 
our theorems on refutation and verification, given in the forthcoming sections,
is that test setups for black-box systems satisfy the
indistinguishability condition.

The indistinguishability condition delimits the knowledge a tester can
obtain through black-box testing.  Suppose Ted (the tester) performs a
black-box analysis of a system $S$. Ted cannot distinguish $S$ from,
say,~$\top$, simply because~$\top$ abstracts every system.  This
epistemic limitation is not alleviated by {\bf complete} tests:
regardless of whether or not Ted samples and analyzes all the
behaviors of~$S$ during testing, $\top\in\ceiling{S}$ is still true.
Rephrasing this in terms of system behaviors, black-box testing can
neither demonstrate the absence of behaviors nor the completeness of
an observation; otherwise, Ted could tell that the system under test
is not~$\top$, which exhibits all behaviors, thereby
distinguishing~$S$ from~$\top$. But, as just discussed, this falls
outside the scope of black-box testing.

\begin{example}
\label{ex:cir0}
Consider Example~\ref{ex:cir}. After observing $t=\{(0,0),(1,1)\}$, the
tester cannot conclude that~$S$ extensionally defines the identity
function. This is not surprising as~$S$ cannot be distinguished
from~$\top$ by observing~$t$ alone, and $\top$ does not define the
identity function. The same argument shows that the tester cannot
conclude that $S$ is deterministic.~\eoe
\end{example}

Note that the indistinguishability condition holds true regardless of
whether or not observations can be carried out in a finite amount of
time; we return to this point in~\S\ref{sec:temporal}. The condition
is also independent from the practical infeasibility of complete tests
(that complete tests are infeasible is demonstrated, e.g.,
in~\cite{kaner}). Moreover, whether the observations are actively
triggered by providing the system under test with selected inputs, or
they are obtained by simply monitoring the system's behaviors is
immaterial; we examine monitoring in~\S\ref{sec:refvsenf}.

We conclude this section with a remark: not all analysis techniques
are constrained by the indistinguishability condition and some,
therefore, can demonstrate the absence of behaviors.  One example is
static analysis, which falls outside the scope of black-box testing as
it relies on a program’s source code as opposed to inspecting a sample
of the program's behaviors~\cite{nielson}.  Similarly, a Fagan
inspection, based on structured reviews of source code and design
documents, is not black-box~\cite{fagan}.  Both of these techniques
can indeed demonstrate the absence of system behaviors.  We return to
the question of how our theory can be extended with additional
information, itself not discernible through black-box testing, in
\S\ref{sec:aux}.  

\section{Refutable Requirements}
\label{sec:refreq}
We formally define the notion of refutability through black-box tests
and prove that any refutable requirement is a prohibition
(\S\ref{sec:refutbbt}). Afterward, we investigate the (ir)refutability
of two important classes of requirements: semi-monotone and
non-semi-monotone requirements (\S\ref{sec:semiref}).

\subsection{Refutability through Black-Box Tests}
\label{sec:refutbbt}

The purpose of testing a system with respect to a requirement is to
refute the hypothesis that the system satisfies the
requirement~\cite{popper,dijkstra70,myers11}. This is in practice
realized by finding a test case where the system does not produce the expected
output.  But for which class of requirements do
such test cases exist?
We characterize below
the class of requirements that can be refuted using black-box tests.
We begin with an illustrative special case that
relates observations, which are refinements of
systems, with requirements, which are sets of systems.

Any system model~$\model=(\Sys,\preceq,\bot,\top)$ induces a {\bf
  reflexive} test setup~$\mathbf{T}_r^\model=(\Sys,\floor{\cdot})$,
where each observation on a system~$S$ is a system in~$\Sys$ that
refines~$S$.
When~$\model$ is clear from the context, we simply
write~$\mathbf{T}_r$ for~$\model$'s reflexive test setup, as we did in
Example~\ref{ex:cir}.  In a reflexive setup, testing a system~$S$
against a requirement~$R$ amounts to inspecting a refinement~$S_w$
of~$S$ to refute the hypothesis~$S\in R$.  By merely observing~$S_w$,
with~$S_w\in \floor{S}$, the tester cannot distinguish~$S$ from any
other system that abstracts~$S_w$, due to the indistinguishability
condition.  Therefore, the tester can infer~$S\not\in R$ after
observing~$S_w$ iff every system in~$\ceiling{S_w}$ violates~$R$.
Hence $R$ is refutable in a reflexive test setup if, for any~$S$ that
violates~$R$, there is at least one {\bf witness}
system~$S_w\in\floor{S}$ such that every system that abstracts~$S_w$
violates~$R$.  That is,~$R$ is refutable in~$\mathbf{T}_r$
if\ $\forall S\in\Sys.\ S\not\in R\to\exists
S_w\in\floor{S}.\ \ceiling{S_w}\cap R=\emptyset$.
\begin{example}
\label{ex:sort}
Consider systems whose input and output domains are the set of lists of
natural number.  Let $R$ be the requirement that
restricts the system's outputs to ascending lists.  Suppose that a
system~$S$ violates~$R$. Then there must exist an input~$i$ for
which~$S$ produces an output list~$o$ that is not ascending. Let us
refer to the system that exhibits just this forbidden behavior
as~$S_w=\{(i,o)\}$. Clearly~$S_w$ refines~$S$, and any system that
abstracts~$S_w$ violates~$R$ by exhibiting the forbidden behavior.
Therefore,~$R$ is refutable in the test setup~$\mathbf{T}_r$.~\eoe
\end{example}

We   generalize the above and define refutability in an arbitrary
test setup.
\begin{definition}
\label{def:refutable}
Let~$\mathbf{T}=(T,\alpha)$ be a test setup for a system
model~$(\Sys,\preceq,\bot,\top)$. A requirement~$R$ is {\bf
  $\mathbf{T}$-refutable} if~$\forall S\in\Sys.\ S\not\in R\to
\exists t\in\alpha(S).\ \hat\alpha(t)\cap R=\emptyset$.
\end{definition}

Let~$R$ be a $(T,\alpha)$-refutable requirement. Then, for any
system~$S$, $S\not\in R\to\ceiling{S}\cap R=\emptyset$, simply
because~$\alpha$ is order-preserving.  The contrapositive implies that
if~$S_1\in R$ and~$S_2\preceq S_1$, then~$S_2\in R$. That is,~$R$ is a
prohibition.  The following theorem is now immediate.

\begin{theorem}
\label{thm:test-neg}
Any $\mathbf{T}$-refutable requirement is a prohibition.
\end{theorem}

We illustrate this theorem with a simple example.
\begin{example}
Consider the model where each system extensionally defines a binary
tree where each node is colored either red or black, and~$\preceq$ is
the subtree relation.  The requirement~$R$ stipulates that the two
children of any red node must have the same color. Observing a
tree~$t$ in which a red node has a red child and a black child implies
that any tree that abstracts~$t$ violates~$R$. Therefore,~$R$ is
refutable in~$\mathbf{T}_r$, and it is a prohibition due to
Theorem~\ref{thm:test-neg}.~\eoe
\end{example}

The following lemma can be seen as a basic sanity check 
on our definition: 
if requirements are refutable, then so is their conjunction.  
\begin{lemma}
\label{lem:compos}
Let $\rho$ be a nonempty set of $\mathbf{T}$-refutable
requirements. Then, $\bigwedge_{R\in\rho}R$ is $\mathbf{T}$-refutable.
\end{lemma}

Given a system model, we say a test setup~$\mathbf{T}_i$ is {\bf more
  permissive} than a test setup~$\mathbf{T}_j$ if any
$\mathbf{T}_j$-refutable requirement is $\mathbf{T}_i$-refutable.  The
following lemma along with Theorem~\ref{thm:test-neg} imply that, in
any system model, the reflexive test setup is the {\bf most
  permissive} test setup.

\begin{lemma}
\label{lem:most-permissive}
In any system model~$\model$, any prohibition
is~$\mathbf{T}_r^\model$-refutable.
\end{lemma}
An intuitive account of this lemma is as follows.  Any test
setup~$\mathbf{T}=(T,\alpha)$ induces a set of
obligations: $\mathcal{O}(\mathbf{T})=\{\hat\alpha(t)\mid t\in T\}$.
Testing a system~$S$ in~$\mathbf{T}$ amounts to the conclusion
that~$S$ satisfies an obligation that includes~$S$, namely the
obligation~$\hat\alpha(t)$, where~$t\in \alpha(S)$ is the observation
obtained through testing.  Therefore, the smaller~$\hat\alpha(t)$ is,
the more we learn about~$S$ by observing~$t$; recall the
indistinguishability condition.  For any system~$S$, the smallest
obligation in~$\mathcal{R}$ that includes $S$ is~$\ceiling{S}$, which
belongs to~$\mathcal{O}(\mathbf{T}_r)=\{\ceiling{S}\mid
S\in\Sys\}$.

We illustrate Lemma~\ref{lem:most-permissive} with an example from
temporal requirements.

\begin{example}
\label{ex:tmpdef}
To investigate temporal requirements, we model systems that induce
infinitely long sequences of events, such as operating systems, and
their requirements following~\cite{DBLP:journals/jcs/ClarksonS10}.
Let~$\Sigma$ be an alphabet, e.g.\ of events or states.  We write
$\Sigma^\omega$ for the set of countably infinite sequences
of~$\Sigma$'s elements.  A behavior is an element of~$\Sigma^\omega$
and a system is a set of behaviors.  The complete
lattice~$(2^{\Sigma^\omega},\subseteq,\emptyset,\Sigma^\omega)$
instantiates our system model, defined in~\S\ref{sec:refut}.

A temporal property~$\phi$ is a set of behaviors. By overloading the
notion of satisfaction, we say a system~$S$ satisfies~$\phi$ if
$S\subseteq \phi$. That is, $\phi$ defines a refinement-closed
requirement: $R_\phi=\floor{\phi}$. Therefore, any property is a
prohibition, hence refutable in $\mathbf{T}_r$ due to
Lemma~\ref{lem:most-permissive}.~\eoe
\end{example}
We return to temporal requirements in~\S\ref{sec:temporal}, where we
show that $\mathbf{T}_r$ can be ``too permissive'' in some settings,
going beyond what is refutable in finite time. 

As~$\mathbf{T}_r$ is the most permissive test setup, a
requirement that is irrefutable in~$\mathbf{T}_r$ is also irrefutable for
any test setup. Obligations are prominent examples of such irrefutable
requirements, as stated in the following lemma. The proof is immediate
by Lemma~\ref{lem:distinct} and Theorem~\ref{thm:test-neg}.
\begin{lemma}
\label{thm:pos-not-ref}
  Nontrivial obligations are irrefutable in any test setup.
\end{lemma}
We illustrate this lemma with an example.

\begin{example}
\label{ex:precirc}
In the $\eio{}$ model, the obligation~$O$ stipulates that systems must
exhibit the behavior~$(1,0)$. Suppose Ted observes~$t=\{(1,1)\}$ while
testing a system~$S$.  Based on $t$, he cannot refute the
hypothesis~$S\in O$, simply because~$\top$ also yields~$t$, and
$\top\in O$.  Of course interpreting~$O$ as the requirement~$P$
stating that \emph{the system may output nothing but~0 for input~1}
results in a refutable requirement. But~$O$ and~$P$ are not
equivalent:~$O$ is an obligation and~$P$ is a prohibition; recall
Lemma~\ref{lem:distinct}.

Note that if it were known (through means outside black-box analysis)
that $S$ is deterministic, then observing~$t$ would justify the
conclusion~$S\not\in O$. We explicate the role of determinacy
assumptions in testing in~\S\ref{sec:practice}.~\eoe
\end{example}

We conclude this section with an intuitive account of (ir)refutability
based on Figure~\ref{fig:pos-neg-ex}. Recall that the figure depicts a
system $S$'s set of behaviors, and its obligation and prohibition,
which are both violated by $S$. To refute the prohibition's
satisfaction, one must locate the black circle in the figure. This is
achievable through black-box testing, which amounts to inspecting a
portion of the triangle (standing for $S$'s set of behaviors). To
refute the obligation's satisfaction, one must locate the white
circle, which lies outside the triangle. This is not achievable
through black-box tests because observations come only from the
triangle's interior.  

Next, we turn to the irrefutability of semi-monotone requirements.

\subsection{The Irrefutability of Semi-Monotone Requirements}
\label{sec:semiref}
Every requirement~$R$ is either (1) semi-monotone, or (2) 
not semi-monotone. In case (1), although~$R$ is irrefutable by
Theorem~\ref{thm:test-neg}, the violation of~$R$ can be demonstrated
through tests for \emph{some} systems. Recall Lemma~\ref{lem:semi}:
$R= \ceiling{R}\wedge \floor{R}$ for a semi-monotone $R$.  Any
system~$S$ that violates~$\floor{R}$ violates~$R$ as
well, and black-box tests can demonstrate $S\not\in\floor{R}$.

\begin{example}
The non-monotone requirement~$R=O\wedge P$, defined in
Example~\ref{eio-def-ex}, is semi-monotone. It states that systems
must extensionally define a total function. 
The system~$S=\{(0,n)\mid n\in\nat\}$ violates $P$, which states that
systems must be deterministic. Any observation that
demonstrates~$S\not\in P$ also demonstrates $S\not\in R$.  Examples
include the observation $\{(0,1),(0,2)\}$ in $\mathbf{T}_r$.  That is,
the hypothesis $S\in R$ can be refuted using  tests that
refute $S\in P$.

Note the contrast to system $S'=\{(0,0)\}$, which violates $R$ but
satisfies $P$. No black-box test refutes~$S'\in R$, simply because it
would have to refute $S'\in O$, contradicting
Lemma~\ref{thm:pos-not-ref}. Recall that $O$ states that systems must
be total.~\eoe
\end{example}

In case (2), where~$R$ is not semi-monotone, it is possible that
testing cannot demonstrate $R$'s violation for \emph{any} system, as the
following example illustrates.
\begin{example}
\label{ex:infflow}
Consider Example~\ref{ex:zigzag}, and the requirement~$R$ defined
there: for each $(i,o)\in S$ there exists some $(i', o)\in S$,
with~$i\neq i'$. Let~$(T,\alpha)$ be a test setup. 
Any observation~$t\in T$ obtained
by testing any system belongs to~$\alpha(\top)$, and~$\top\in R$. That
is, through black-box tests, we cannot distinguish any system
from~$\top$, which indeed satisfies~$R$. Therefore,~$R$'s violation
(for any system) cannot be demonstrated through tests in any test
setup.~\eoe
\end{example}

To summarize, a requirement is refutable through black-box tests iff
it is a prohibition. Nontrivial obligations are irrefutable, and so
are non-monotone requirements. However, the violation of a
semi-monotone requirements that is not monotone can be demonstrated
through tests, but only for some of the systems that violate them. It
is possible that the violation of the requirements that are not
semi-monotone cannot be demonstrated through black-box tests for any
system.\footnote{The reason for ``it is possible'' is that a
  requirement that is not semi-monotone can have a semi-monotone
  component. For instance, let~$R$ be a requirement that is
  not semi-monotone and $\top\not\in R$. Define the prohibition
  $P=\Sys\setminus \{\top\}$. Note that $R \wedge P=R$. Clearly
  $\top\not\in P$ can be demonstrated through tests, thereby refuting
  the hypothesis $\top\in R$.}

\section{Verifiable Requirements}
\label{sec:verif}
We define testing with the purpose of verifying the satisfaction of a
requirement as dual to testing for refutation.
\begin{definition}
\label{def:verif}
Let~$\mathbf{T}=(T,\alpha)$ be a test setup for a system
model~$(\Sys,\preceq,\bot,\top)$. A requirement~$R$ is {\bf
  $\mathbf{T}$-verifiable} if~$\forall S\in\Sys.\ S\in R\to
\exists t\in\alpha(S).\ \hat\alpha(t)\subseteq R$.
\end{definition}

In particular, a requirement~$R$ is $\mathbf{T}_r^\model$-verifiable
in the system model~$\model=(\Sys,\preceq,\bot,\top)$ if\ $\forall
S\in\Sys.\ S\in R \to \exists S_w\in\floor{S}.\ \ceiling{S_w}\subseteq
R$. That is, if there exists a witness system~$S_w$ that refines~$S$
and any system that abstracts~$S_w$ satisfies~$R$, then by
observing~$S_w$ we have conclusively demonstrated~$S\in R$.
The following
theorem is dual to Theorem~\ref{thm:test-neg}.
\begin{theorem}
\label{thm:dual}
Any $\mathbf{T}$-verifiable requirement is an obligation.
\end{theorem}

An observation~$t\in\alpha(S)$ proves that the system~$S$ satisfies
the obligation $O=\hat\alpha(t)$. It also proves that $S\in R$ for any
requirement~$R\supseteq O$. Therefore, as $O$ becomes smaller, more
requirements are proved by the observation.
This explains why~$\mathbf{T}_r$ is the most permissive test setup for
verification: any~$\mathbf{T}$-verifiable requirement
is~$\mathbf{T}_r$-verifiable.
Consequently, a requirement that is not~$\mathbf{T}_r$-verifiable is
non-verifiable in any test setup.
Prohibitions are prominent examples of such non-verifiable
requirements, as the following lemma states.
\begin{lemma}
\label{lem:prohver}
Nontrivial prohibitions are non-verifiable in any test setup.
\end{lemma}  
It is instructive to compare this lemma and Dijkstra's often-quoted
statement~\cite{dijkstra70} in the context of black-box testing
that ``program testing can be used to show the presence of bugs, but
never to show their absence.''
 Dijkstra argues that programs have large,
typically infinite, input domains. It is intractable to test a
program's behavior for each input. It follows then that testing
cannot prove the absence of bugs, i.e.\ deviations from expected
behavior. Note the contrast to Lemma~\ref{lem:prohver}, which holds
even if a tester could run an infinite number of tests. As
discussed in~\S\ref{sec:bbt}, non-determinism poses an epistemic
limitation on what testing can achieve, regardless of the cardinality
of input domains and the number of tests we execute.

Contrary to the folklore, Lemma~\ref{lem:prohver}, and by extension
Dijkstra's statement, do not imply that no requirements are verifiable
through black-box tests.
For instance, the requirement that obliges a magic 8-ball to output
\emph{ask again later} is clearly verifiable through  tests:
observing this output once demonstrates the obligation's satisfaction.
We further illustrate this point with an example from temporal
requirements.

\begin{example}
\label{ex:ctl}
Consider the system model of Example~\ref{ex:tmpdef}. Let~$e$ be an
element of $\Sigma$. The requirement~$R_e$ consists of the systems
that exhibit at least one behavior in which~$e$ appears.
Note that~$R_e$ is an obligation since its satisfaction is
abstraction-closed.  Let $S \in R_e$. Observing a refinement~$S_w$
of~$S$ where~$S_w$ exhibits one behavior~$\pi$ in which~$e$ appears
demonstrates~$S\in R_e$: any abstraction of~$S_w$ exhibits~$\pi$ as
well, hence satisfying~$R_e$. We conclude that the obligation~$R_e$ is
verifiable through tests in~$\mathbf{T}_r$.~\eoe
\end{example}

We can now sharpen {\bf Dijkstra's dictum} to: 
\begin{quote}
 {\bf (D)}\quad Program testing can be used to show the presence of
 behaviors, but never to show their absence, even if an infinite
 number of tests were allowed.
\end{quote}
If a software \emph{bug} is a prohibited behavior, then {\bf (D)}
extends Dijkstra's statement, simply stipulating that prohibitions are
refutable, but not verifiable. However, if a \emph{bug} is the absence
of an obliged behavior, then {\bf (D)} translates to: program testing
can be used to show the absence of bugs, but never to show their
presence.  This sentence unrolls to: program testing can be
used to show the presence of obliged behaviors, but never to show
their absence.  In other words, obligations are verifiable, but not
refutable. That tests cannot show the absence of obliged behaviors has
tangible implications. For example, fuzz testing can hardly reveal
omission bugs, i.e.\ bugs due to developers' failure to implement a
desired feature or functionality~\cite{fuzzing-book}.  

The folklore that testing is capable of refutation, but not proving
correctness, is sometimes held by members of the software testing community
and presumably reflects the wide-spread testing of prohibitions.
An example of this is the statement in \cite{Aichernig:2007}:
    ``Rather than doing verification by testing, a doubtful
     endeavour anyway, here we focus on falsification. It is falsification,
     because the tester gains confidence in a system by designing test cases
     that would uncover an anticipated error. If the falsification fails, it
     follows that a certain fault does not exist.''
But testing is not only a refutation technique: it is also a proof
technique, as it can prove that the system under test satisfies an
obligation, such as the one given in Example~\ref{ex:ctl}.

\section{Refutation in Finite Time}
\label{sec:temporal}
A requirement that is deemed refutable in our theory might not be
refutable in practice. For example, a requirement whose refutation
hinges upon measuring the exact momentum and position of a quantum
object is impossible to refute due to the laws of physics.  This
limitation, not unexpectedly, does not follow from our  theory
of black-box tests.
Below, we extend our theory to account for a
practically relevant limitation of system testing: we consider
refutation through black-box tests that proceed in a finite amount of
time. We define the notion of finite refutability (\S\ref{sec:finref}), and
illustrate it by characterizing finitely refutable temporal properties
and hyper-properties (\S\ref{sec:hyper}).

\subsection{Finite Refutability}
\label{sec:finref}
Intuitively, a test proceeds in a finite amount of time if
observations can be carried out in finite time. 
This motivates the following definition.

\begin{definition}
\label{def:finref}
Let~$\mathbf{T}=(T,\alpha)$ be a test setup for a system model
$(\Sys,\preceq,\bot,\top)$. A requirement~$R$ is {\bf finitely
  refutable} in $\mathbf{T}$ if 
\begin{enumerate}[(i)]
\item $R$ is $\mathbf{T}$-refutable, and
\item every observation in~$\bigcup_{S\in\Sys}\alpha(S)$ can be
  carried out in finite time.
\end{enumerate}
The notion of {\bf finite verifiability} is defined dually.
\end{definition}

Condition (ii) of Definition~\ref{def:finref} refers to the world:
determining whether an arbitrary observation can be carried out in
finite time falls outside our theory's scope, and this condition's
satisfaction must be substantiated by other means. Thus our theory
cannot establish a requirement's finite (ir)refutability unless
assumptions are made about what can be observed in finite time in the
world.

To illustrate Condition (ii), we consider a family of test setups for the
$\eio{}$ system model; a similar notion can be defined for other
system models. The {\bf family of test
  setups}~$\mathbf{T}_k=((\nat\times\nat)^k,\alpha_k)$, with~$k\ge 1$,
and $\alpha_k$ maps system~$S$ to~$S^k$ and is inductively
defined by~$S^1=S$ and~$S^{k+1}=S\times S^{k}$.  Clearly $\alpha_k$ is
order-preserving for any $k \in\nat$, and testing a system~$S$ in the
setup~$\mathbf{T}_k$ amounts to observing~$k$ input-output pairs
belonging to~$S$.   We can now illustrate Condition (ii)  as follows.
\begin{example}
\label{ex:tn}
Consider the $\eio{}$ system model, and assume that natural numbers
are observable in finite time. Then, observing any element
of~$(\nat\times\nat)^k$, where~$k\ge 1$ belongs to~$\nat$, takes
finite time. The requirement~$R_\nz$ stating that \emph{systems never
  output zero} is, under this assumption, finitely refutable
in~$\mathbf{T}_1$.
Now consider the requirement~$R_\fz$, stating that \emph{systems may
  output zero for at most finitely many inputs}.  Note that~$R_\fz$ is
refutable in the reflexive test setup~$\mathbf{T}_r$,  but not
finitely so, and moreover it is
not~$\mathbf{T}_k$-refutable for any~$k\ge 1$.

It now seems reasonable to conclude that~$R_\fz$ is not finitely
refutable: no finite set of behaviors can refute~$R_\fz$.  This
conclusion does not however follow from our theory. To illustrate,
consider an alternative test setup~$\mathbf{T}=(\{*,\times\},
\alpha)$, where~$\alpha(S)=\{*\}$ if~$S$ outputs zero for finitely
many inputs, and~$\alpha(S)=\{*,\times\}$ otherwise. Since~$\alpha$ is
order-preserving,~$\mathbf{T}$ is formally a test setup.  The
requirement~$R_\fz$ is finitely refutable in~$\mathbf{T}$, under the
assumption that $\alpha(S)$ is observable in
finite time. Whether this is a tenable assumption cannot be settled
inside our theory.  Although~$\mathbf{T}$ hardly appears realizable,
such observations are possible in certain cases, for example by
measuring the electromagnetic radiation emitted from a black-box
system; see, e.g.,~\cite{Standaert:2009:UFA:1533674.1533709}.~\eoe
\end{example}

Although the satisfaction of Condition~(ii) cannot be
settled in our theory, this condition has implications
relevant to our theory: any test setup in which observations amount to
inspecting infinite objects cannot be used to show finite
refutability. For example, it follows that only finite portions of
finitely many system behaviors can be inspected for each observation,
even if those behaviors are not themselves finite objects.  We
illustrate this point with an example.  
\begin{example}
\label{ex:real}
Consider the system model~$(2^{\mathbb{R}\times\mathbb{R}}, \subseteq,
\emptyset, \mathbb{R}\times\mathbb{R})$, where~$\mathbb{R}$ is the set
of real numbers. This system model is similar to the~$\eio{}$ model
except its inputs and outputs are real numbers.

Define~$\prefix(r)$ as the set of finite truncations of the decimal
expansion of a real number~$r$. For
instance,~$\prefix(\sqrt{2})=\{1,1.4,1.41,1.414,\cdots\}$. Note that a
real number can have more than one decimal expansions, for example, 1
and $0.999\cdots$, but accounting for this point is unnecessary for
our discussion here. 

We define the test setup~$\mathbf{T}=(\mathbb{F}\times\mathbb{F},
\alpha)$, where~$\mathbb{F}$ is the set of rational numbers that have
a finite decimal expansion and~$\alpha$ maps any system~$S$ to the
set~$\bigcup_{(i,o)\in S}\prefix(i)\times\prefix(o)$.  An observation
of a system $S$ in this setup is a pair $(f_1, f_2)$, where $f_1$ is a
truncation of an input $i$ and $f_2$ is a truncation of an output $o$,
where $(i, o) \in S$. That is, we may observe only finite portions of
the decimal expansions of the inputs and outputs. If~$\mathbb{F}$'s
elements are observable in finite time, then any
$\mathbf{T}$-refutable requirement is finitely refutable.

Now consider the requirement $R_<$, stating that system outputs are
strictly smaller than $\sqrt 2$. Clearly $R_<$ is a prohibition, hence
$\mathbf{T}_r$-refutable.  Define the system $S=\{(1,
1.4142\cdots)\}$, which outputs $\sqrt 2$, decimally expanded, for the
input 1. Even though $S$ violates $R_<$, no truncation of $S$'s
output's decimal expansion conclusively demonstrates this, because the
set of permissible outputs according to~$R_<$, namely $\{o
\in\mathbb{R}\mid o < \sqrt 2\}$, is not a closed set
in~$\mathbb{R}$'s standard topology. That is, there is a number,
namely $\sqrt 2$, that is arbitrarily close to the elements of this
set, but is not a member of the set. No finite truncation of this
number's decimal expansion can therefore conclusively determine
whether it is a member, or not. We conclude that $R_<$ is not
$\mathbf{T}$-refutable.

An analogous argument shows that the requirement $R_\le$ stating that
system outputs must be less than or equal to $\sqrt 2$ is
$\mathbf{T}$-refutable, and hence finitely refutable, because the set
of permissible outputs it induces, namely~$(-\infty,\sqrt{2}]$, is
  topologically closed.~\eoe
\end{example}

The example hints at a fundamental connection between refutability and
topological closure when system behaviors are infinite sequences. This
connection was investigated by Alpern and Schneider in the
context of temporal properties~\cite{AlpernS85}, which we turn to next.

\subsection{Finitely Refutable Temporal Requirements}
\label{sec:hyper}
In this section, we characterize finitely refutable temporal
requirements.  We start by extending the system
model~$(2^{\Sigma^\omega},\subseteq,\emptyset,\Sigma^\omega)$,
associated with temporal requirements (defined in
Example~\ref{ex:tmpdef}), with some additional temporal notions.

Let $\Sigma^*$ be the set of all finite sequences of $\Sigma$'s
elements.  For a behavior~$\pi\in \Sigma^\omega$, we
write~$\prefix(\pi)$ for the set of all finite prefixes of~$\pi$, and
define the test setup~$\mathbf{T}_*$ as~$(T_*,\alpha_*)$, where~$T_*$
is the set of all finite subsets of~$\Sigma^*$, and~$\alpha_*(S)$ is
the set of all finite subsets of~$\bigcup_{\pi\in S} \prefix(\pi)$ for
a system~$S$.
Intuitively, any element of~$\alpha_*(S)$ is a possible observation
of~$S$, where finite prefixes of finitely many behaviors of~$S$ are
inspected.  For any~$\mathbf{T}_*$-refutable requirement~$R$ and
any~$S\not \in R$, there exists a finite (witness) set~$t_w$ of finite
prefixes of~$S$'s behaviors such that any system~$S'$ that could have
yielded the observation~$t_w$, i.e.\ $t_w\in \alpha_*(S')$,
violates~$R$.  Clearly, any~$\mathbf{T}_*$-refutable requirement is
finitely refutable, if elements of~$\Sigma$ can be observed in finite
time.  Next, we relate~$\mathbf{T}_*$-refutability
and~$\mathbf{T}_*$-verifiability to the notions of temporal properties
and hyper-properties.

A temporal {\bf property} is a set of permissible
behaviors~\cite{pnueli,AlpernS85}, i.e.\ a subset of~$\Sigma^\omega$.
Any property~$\phi$ defines a prohibition, namely the
refinement-closed requirement $R_\phi=\floor{\phi}$ (recall
Example~\ref{ex:tmpdef}).
We illustrate this with an example.
\begin{example}
\label{ex:ltlvsctl}
Recall the requirement~$R_e$ from Example~\ref{ex:ctl}, consisting of
the systems that exhibit at least one behavior where~$e$ appears.
Since properties are prohibitions, this obligation
is not a property as otherwise~$R_e$ would have to
be trivial by Lemma~\ref{lem:distinct}.

As a side note, that~$R_e$ is not a property shows that $R_e$ cannot
be expressed as a formula in the linear-time temporal logic (LTL),
whose formulas define properties~\cite{pnueli}.  Since $R_e$ is
expressed as~$\mathsf{EF}\ e$ in the computation tree logic CTL, we
can conclude the well-known result that LTL is \emph{not} more
expressive than CTL; for an introduction to CTL and its expressiveness
see~\cite{ctl-star}.~\eoe
\end{example}

We now turn to safety and liveness property types.  We denote the
concatenation of an element of $\Sigma^*$ with one of $\Sigma^\omega$
by their juxtaposition. A property~$\phi$ is {\bf safety} if~$\forall
\pi\not\in \phi.\,\exists \sigma\in \prefix(\pi).\, \forall
\pi'\in\Sigma^\omega.\ \sigma\pi'\not\in \phi$ and {\bf liveness}
if~$\forall \sigma\in\Sigma^*.\exists \pi\in\Sigma^\omega.\ \sigma
\pi\in \phi$. That is, safety and liveness properties are closed and
dense sets, respectively~\cite{AlpernS85}.
The following lemma, whose proof hinges upon Lemma~\ref{lem:distinct}
and Theorem~\ref{thm:test-neg}, implies that nontrivial liveness
properties, although $\mathbf{T}_r$-refutable, are
not~$\mathbf{T}_*$-refutable; cf.~\cite{AlpernS85}.

\begin{lemma}
\label{lem:temporal}
A temporal property~$\phi$ is~$\mathbf{T}_*$-refutable iff~$\phi$ is
safety. Moreover, all temporal properties are
$\mathbf{T}_r$-refutable and any $\mathbf{T}_*$-verifiable property
is trivial.
\end{lemma}

Any property~$\phi$ is the conjunction of a safety property~$\phi_s$
and a liveness property $\phi_l$~\cite{AlpernS85}. Therefore, if a
system~$S$ violates $\phi_s$, then the hypothesis $S\in\phi$ can be
refuted through tests aimed at refuting $S\in \phi_s$. However,
if~$S$ violates only the liveness conjunct of $\phi$, namely $S\in
\phi_s$ and $S\not\in \phi_l$, then the hypothesis $S\in \phi$ cannot
be (finitely) refuted in $\mathbf{T}_*$, due to
Lemma~\ref{lem:temporal}. This is akin to the refutability of
semi-monotone requirements for \emph{some} systems, discussed
in~\S\ref{sec:refreq}. See also~\cite{approx,more}.

We now turn to hyper-properties.  A {\bf hyper-property} is a set of
properties~\cite{DBLP:journals/jcs/ClarksonS10}, i.e.\ a requirement in
our model. A system~$S$ satisfies a hyper-property~$\mathbb{H}$,
if~$S\in \mathbb{H}$.  In our setting, a hyper-property~$\mathbb{H}$ is
{\bf hyper-safety} \cite{DBLP:journals/jcs/ClarksonS10} if for
any~$S\not\in \mathbb{H}$, there exists an observation~$t\in \alpha(S)$
such that~$\forall S'\in\hat\alpha(t).\ S'\not\in \mathbb{H}$; It is
easy to check that a temporal requirement~$R$ is hyper-safety iff~$R$
is~$\mathbf{T}_*$-refutable.
Now it is immediate by Theorem~\ref{thm:test-neg} that any
hyper-safety requirement is a prohibition. Therefore, finitely
verifiable hyper-safety requirements must be trivial.
For instance, Example~\ref{ex:ctl}'s requirement~$R_e$, which is
clearly finitely verifiable in~$\mathbf{T}_*$, cannot be hyper-safety
and it is therefore not finitely refutable in~$\mathbf{T}_*$.  
These results show how existing, specialized concepts and their
refutability follow as special cases of the notions we defined.
We revisit properties and hyper-properties in~\S\ref{sec:refvsenf}.

In light of this discussion, one
  can see the test setup~$\mathbf{T}_k$, defined
  in~\S\ref{sec:finref}, as the setup where a single observation is
  carried out over $k$ copies of the system under test. As a side remark,
  we note that for each $k$ there is a requirement that is
  $\mathbf{T}_{k+1}$-refutable but not $\mathbf{T}_k$-refutable. That
  is, self-composing a system $k$ times, namely observing $k$ system
  behaviors, is not sufficient for demonstrating that the system
  violates the requirement. For example, sampling a curve three times
  is not sufficient for refuting the requirement stating that the
  curve is a circle: there is a circle that passes any three points on
  the plane. Sampling the curve four times could however refute this
  particular requirement.

\section{Refutability through Algorithmic Means}
\label{sec:alg}

A {\bf test oracle} for a requirement $R$ is a (partial) decision
function that given an observation on a system decides whether the
system violates $R$.  Our definition of finitely refutable
requirements poses no constraints on their test oracles. In
particular, whether an observation demonstrates a violation of a
finitely refutable requirement need not be decidable. Such
``undecidable'' requirements are uncommon in testing
practice. However, we show that explicating the computational
constraints of refutation not only clarifies the limits of algorithmic
testing (\S\ref{sec:algref}),  it also sheds light on the
relationship between refutation through testing and enforcement
through monitoring (\S\ref{sec:refvsenf}).

\subsection{Algorithmic Refutability}
\label{sec:algref}
We start with two auxiliary definitions, and assume that the reader is
familiar with basic computability theory. For an introduction to this topic
see, e.g.,~\cite{rogers}.

Given a countable set~$U$, a set~$E\subseteq U$ is {\bf recursively
  enumerable} if there is a (semi-)algorithm~$\mathcal{A}_E$ that
terminates and outputs \emph{true} for any input~$u\in U$ that is a
member of~$E$. If~$u\not\in E$, then~$\mathcal{A}_E$ does not
terminate.

Any requirement~$R$ induces a set~$\Omega_R$ of {\bf irremediable
  observations}~$\{t\in T\mid \hat\alpha(t)\cap R=\emptyset\}$ in a
test setup~$\mathbf{T}=(T,\alpha)$.
It follows that a system~$S$ violates a~$\mathbf{T}$-refutable~$R$
iff~$\alpha(S)\cap \Omega_R\neq\emptyset$.  
Intuitively, a requirement is algorithmically refutable 
 if it induces
a recursively enumerable set of irremediable
observations.
 This is
because if a system~$S$ violates such a requirement~$R$
in~$\mathbf{T}$, then there is at least one observation~$t\in
\alpha(S)$ that can be made in finite time,
where~$\mathcal{A}_{\Omega_R}$ terminates on~$t$ and outputs
\emph{true}. Here, \emph{true} means~$t\in\Omega_R$,
demonstrating~$S\not\in R$. Observing such a~$t$ through testing,
therefore, conclusively refutes the hypothesis~$S\in R$.
\begin{definition}
\label{def:alg}
Let~$\mathbf{T}=(T,\alpha)$ be a test setup for a system model
$(\Sys,\preceq,\bot,\top)$.  A requirement~$R$ is {\bf algorithmically
  refutable} in $\mathbf{T}$ if~$R$ is finitely refutable
in~$\mathbf{T}$, and~$\Omega_R$ is a recursively enumerable subset of
the countable set~$T$.
\end{definition} 
\noindent
Note that, from the standpoint of refutation, nothing is
gained by requiring the set $\Omega_R$ to be recursive
since determining that an element is not an irremediable observations does not contribute to
  the requirement's refutation.  It is therefore not necessary to
  determine non-membership. 

The following example illustrates Definition~\ref{def:alg}.

\begin{example}
In $\eio{}$, the prohibition~$P$ states that a system may never
output~$0$ for an odd input. Clearly, $P$ is~$\mathbf{T}_1$-refutable,
and its set of irremediable observations~$\Omega_P=\{\{(2i+1,0)\}\in
T_1\mid i\in\nat\}$ is recursive.
If natural numbers are observable in finite time, then~$P$ is
algorithmically refutable in~$\mathbf{T}_1$: any~$S$ that violates~$P$
induces an observation, for example~$t=\{(3,0)\}$, where a Turing machine
arrives at the verdict~$t\in \Omega_P$ in finite time.~\eoe
\end{example}

Let~$R$ be an algorithmically refutable requirement
in~$\mathbf{T}=(T,\alpha)$, and~$S$ be a system where~$\alpha(S)$ is a
recursively enumerable subset of~$T$.  Note that, as previously
discussed, black-box testing cannot establish the absence of behaviors
in the system under test~$S$. Therefore, it is crucial that $\alpha(S)$
is recursively enumerable, but not, say, co-recursively enumerable or
recursive: these would imply that the tester could ``see'' absent
behaviors.  The decision problem that asks whether~$S$ violates~$R$ is
semi-decidable, as the following test algorithm illustrates.

\begin{algorithm}[Test Algorithm]
\label{alg}
Fix an arbitrary total order on~$T$'s elements.
Dovetail~$\mathcal{A}_{\alpha(S)}$'s computations on the elements
of~$T$.\footnote{Dovetailing, which is a primitive parallelization
  technique, proceeds in stages. Given an ordered list of inputs
  $x_0,x_1,\cdots$, one step of computation is performed on $x_0$ in
  stage 1. In stage $n+1$, we perform $n+1$ steps of the computations
  for $x_0, \cdots,x_n$. In contrast to performing the computations on
  $x_0$, and then on $x_1$, and so forth, the benefit of dovetailing
  is the following: suppose the computation for $x_0$ never
  terminates, whereas it terminates for $x_1$. Then, dovetailing's 
  parallelization ensures that in a finite amount of time the result
  of the computation on $x_1$ becomes available. 

As a side note, dovetailing a system $S$'s executions is not hindered by
$S$ being a black-box. To perform dovetailing, a tester merely needs a
mechanism for controlling the progress of $S$'s computations.  For
example, when $S$ is given as a computer program, dovetailing can be
achieved by controlling the CPU cycles allocated to $S$'s
computations, which does not require inspecting the program's source
code.}  In parallel, dovetail~$\mathcal{A}_{\Omega_R}$'s computations
on those observations for which~$\mathcal{A}_{\alpha(S)}$
terminates. Output \emph{true} and terminate, when a computation
of~$\mathcal{A}_{\Omega_R}$ terminates.
\end{algorithm}

The algorithm checks whether the intersection of two recursively
enumerable sets, namely $\alpha(S)$ and $\Omega_R$, is empty.
If~$S\not \in R$, then there exists at least one observation~$t_w$ in
the set~$\alpha(S)\cap\Omega_R$. The test algorithm is bound to
terminate on~$t_w$ and output \emph{true}, thus
demonstrating~$S\not\in R$ in finite time.  However, if~$S\in R$, then
the test (semi-)algorithm does not terminate.

Algorithm~\ref{alg} achieves the (impractical) ideal of testing: it
not only has ``a high probability of detecting an as yet undiscovered
error''~\cite{myers11}, the algorithm is in fact guaranteed to reveal
flaws in any system that violates a requirement, if the preconditions
are met. In this sense, Algorithm~\ref{alg} demonstrates the limits of
algorithmic testing. We remark that although this algorithm is not a
recipe for testing practice, there are similarities. For instance,
standard test selection methods place likely witnesses of~$R$'s
violation early in the ordering assumed on~$T$~\cite{myers11}. This
would speed up Algorithm~\ref{alg} too.

We illustrate Algorithm~\ref{alg} with an example.
\begin{example}
Let $M$ be a Turing machine that is available to us as a black-box: we
may provide $M$ with an input $i$ and observe \emph{halt} if $M$ halts
on $i$. If $M$ diverges on $i$, then we observe no outputs.

In the $\eio{}$ model, the requirement~$R$ is defined as:~$S\in R$ if
for any~$(i,1)\in S$ the machine~$M$ diverges on the
input~$i$. It is easy to check that~$R$ is
$\mathbf{T}_1$-refutable, with $\mathbf{T}_1=(T_1,\alpha_1)$, and the
set~$\Omega_R$ is recursively enumerable: given an observation
$\{(i,1)\}$ in $T_1$, if $\{(i,1)\}\in \Omega_R$, then $M$ is bound to
halt on $i$.

Suppose that~$\alpha_1(S)$ is recursively enumerable for a
system~$S$. That is,~$\mathcal{A}_{\alpha(S)}$ is guaranteed to
terminate on any~$(i,o)\in S$.  If~$S\not\in R$, then there is a
witness~$(w,1)\in S$ where~$M$ terminates on~$w$. Therefore,
dovetailing~$M$'s computations on~$i$ for all~$(i,1)$ on
which~$\mathcal{A}_{\alpha(S)}$ terminates is bound to exhibit a
terminating computation, thus demonstrating~$S\not\in R$ in finite
time.

Now consider the requirement~$U$ stating that a system may contain
$(i,1)$ only if $M$ halts on $i$. Clearly~$U$ is
$\mathbf{T}_1$-refutable. However,~$\Omega_U=\{\{(i,1)\}\mid i\in\nat,
M\ \mbox{diverges on}\ i\}$ is not recursively enumerable: $M$
produces no outputs for such an input $i$. This shows that $U$ cannot
be algorithmically refuted in $\mathbf{T}_1$.~\eoe
\end{example}

Next we apply the notion of algorithmic refutability to establish a
duality between refutation and enforcement in the context of temporal
requirements.

\subsection{Refutation versus Enforcement}
\label{sec:refvsenf}
It is easy to check that a safety property~$\phi$ is algorithmically
refutable in $\mathbf{T}_*$ iff~$\phi$'s set of {\bf irremediable
  sequences}~$\nabla_\phi=\{\sigma\in \Sigma^*\mid \forall
\pi\in\Sigma^\omega.\ \sigma\pi\not\in \phi\}$ is recursively
enumerable.
This condition separates refutability from {\bf enforceability}, as
explained below. To enforce the safety property~$\phi$ on a
system~$S$, a reference monitor observes some~$t\in \alpha(S)$. If~$t$
demonstrates that~$S$ violates~$\phi$, then the monitor
\emph{stops}~$S$. Otherwise, the monitor \emph{permits}~$S$ to
continue its execution. For enforcement, the set~$\nabla_\phi$ must
therefore be recursive~\cite{Hamlen}.
It then follows that any enforceable temporal property is
algorithmically refutable. In contrast, any property~$\phi$, where~$\nabla_\phi$
is recursively enumerable but not recursive, is algorithmically
refutable, but not enforceable.

To further illustrate the relationship between refutability and
enforceability, we define {\bf weak enforceability} for a hyper-safety
requirement~$R$ as follows.  By monitoring the executions of a
system~$S$, a monitor observes some~$t\in\alpha_*(S)$. If~$t$ does
\emph{not} conclusively demonstrate~$S\not\in R$,
then the monitor permits~$S$ to continue. However, if~$t$ does
conclusively demonstrate~$R$'s violation, then the monitor
may either stop~$S$, or diverge and thereby stall~$S$.
Recall that a system~$S$ violates a hyper-safety requirement~$R$
iff~$\alpha_*(S)\cap \Omega_{R}\neq\emptyset$.
To weakly enforce~$R$, the set~$\Omega_R$ must therefore be
co-recursively enumerable, i.e.~$T_*\setminus\Omega_R$ must be
recursively enumerable.  This observation, which
concurs with~\cite[Theorem 4.2]{popl15}, illustrates that weak
enforceability and algorithmic refutability are complementary in the
sense that the former requires~$\Omega_R$ to be co-recursively
enumerable and the latter requires~$\Omega_R$ to be recursively
enumerable.
This duality between refutability and enforceability becomes evident
only after explicating the computational constraints of testing and
enforcement.

\section{Refutability under Auxiliary Assumptions}
\label{sec:aux}
Testers might have partial information about a black-box system. For
example, they might have knowledge that an otherwise black-box system
is deterministic. In this section we explore how black-box testing can
be augmented with such information. Strictly speaking, this topic
falls outside the black-box setting of~\S\ref{sec:refut}, where the
  domain~$\Sys$ contains at least one non-determinism system,
  namely~$\top$. Testing under auxiliary assumptions can in fact be
  seen as a form of {\bf gray-box testing}, where testers know,
through means outside of black-box analysis, that a system's internal
operations are in some way constrained.

\subsection{Generalized Refutability}

We start by generalizing the definition of refutability to
\emph{refutability under auxiliary assumptions} since a tester's
partial information about the system under test can be modeled as
assumptions.
We define an {\bf assumption} as a set of systems. An assumption~$A$
is valid for a system~$S$, or $S$ satisfies $A$, if $S$ is included in
$A$. Clearly assumptions and requirements have the same type in our
model. However, in contrast to requirements, assumptions are in
general not subjected to analysis.

An assumption~$A$ about the system under test~$S$ can alleviate a
tester's epistemic limitation (rooted in the indistinguishability
condition).  Namely, if $S$ is assumed to satisfy $A$, then from an observation
$t$ the tester can conclude that $S$ belongs to a subset of
$\hat\alpha(t)$, namely the subset that consists of the systems that
satisfy $A$. This motivates the following definition.

\begin{definition}
\label{def:refutunder}
Let $(\Sys,\preceq,\bot,\top)$ be a system model,
$\mathbf{T}=(T,\alpha)$ a test setup, and $A$ an assumption. A
requirement $R$ is $\mathbf{T}$-{\bf refutable under} $A$ if $\forall
S\in A.\ S\not\in R\ \to\ \exists t\in\alpha(S).\ (\hat\alpha(t)\cap
A)\cap R =\emptyset$.
\end{definition}
Note that letting $A=\Sys$ in Definition~\ref{def:refutunder}
results in Definition~\ref{def:refutable}. That is, refutability is a
special case of refutability under assumptions where no assumptions
are made about systems.
We illustrate Definition~\ref{def:refutunder} with an example.

\begin{example}
\label{ex:orefut}
Consider Example~\ref{ex:precirc}, where the obligation $O$ consists
of the systems that exhibit the behavior $(1,0)$. We define the
assumption $A$ as the set of total, deterministic systems, to reflect
the information that the system under test is known to be total and
deterministic.

The obligation $O$ is $\mathbf{T}_1$-refutable under $A$, simply
because any system that satisfies $A$ and violates $O$ must exhibit a
behavior $(1,i+1)$, with $i\in \nat$, due to totality. Observing such
a behavior demonstrates that the system does not exhibit $(1,0)$, due
to determinacy. Therefore the system violates $O$.

Clearly the above argument falls apart without the determinacy
assumption. To see why $A$'s totality conjunct is also necessary for
the argument, consider the deterministic system $S=\{(0,0)\}$, which
violates $O$ as well as the totality conjunct of $A$. No observation
in $\alpha_1(S)$ can demonstrate $S\not\in O$.~\eoe
\end{example}

Example~\ref{ex:orefut} shows that irrefutable requirements can be refuted if the
system under test is known to satisfy certain assumptions. This is
because an assumption $A$ reduces the set of the systems
that could have yielded a given observation. The smaller $A$ is, the
weaker the indistinguishability condition becomes.  Note however that
the set of test subjects shrinks along with $A$: refutability under
$A$ pertains only to the systems in $A$.

The process of adding assumptions to obtain refutability does not
undermine Theorem~\ref{thm:test-neg}'s statement that prohibitions are
the only requirements that can be refuted through tests.
Theorem~\ref{thm:test-neg} is a special case of the following theorem
where $A=\Sys$: that \emph{any refutable
  requirement is a prohibition} pertains to the special case where no
assumptions are made about systems.
\begin{theorem}
\label{thm:replace} 
Let $\mathbf{T}=(T,\alpha)$ be a test setup, and $A$ an assumption.
If a requirement $R$ is $\mathbf{T}$-refutable under $A$, then there
exists a prohibition $P_{R\mid A}$ such that $\forall S\in
A.\ S\not\in P_{R\mid A} \leftrightarrow S\not\in R$. Namely,
$P_{R\mid A}=\floor{A\cap R}$.
\end{theorem}
\noindent This theorem implies that the task of refuting an (irrefutable) $R$
under $A$ can be reduced to the task of refuting the prohibition
$P_{R\mid A}$.

The following example illustrates
Theorem~\ref{thm:replace}.

\begin{example}
\label{ex:rgivena}
Consider Example~\ref{ex:orefut}. Note that $A\cap O$ consists of
deterministic total systems that exhibit the behavior $(1,0)$.  We
define~$P$ as the requirement that forbids exhibiting any behavior
$(1,i+1)$, with $i\in\nat$. Roughly speaking, $P$ is found by adding
the closed-world assumption~\cite{Reiter} to $O$: the outputs not
obliged by $O$ are prohibited by $P$.
Below, we show $P$ is equal to $P_{O\mid A}$, when confining our
attention to the systems that satisfy $A$.

If a deterministic total system $S$ belongs to $P_{O\mid A}$, then $S$
belongs to $O$ due to Theorem~\ref{thm:replace}. That is, $(1,0)\in
S$. Then, due to $S$'s determinacy, no behavior $(1,i+1)$, with $i\in
\nat$, can belong to $S$. Therefore, $S\in P$.

If a deterministic total system $S$ belongs to $P$, then, due to its
totality and determinacy, $S$ must exhibit $(1,0)$. Therefore, $S\in
O$, and hence $S\in A\cap O$. This entails $S\in \floor{A\cap
  O}$. That is, $S\in P_{O\mid A}$.

That $P = P_{O\mid A}$ further justifies the definition of $P_{R\mid
  A}$, given in Theorem~\ref{thm:replace}: $P_{O\mid A}$ is indeed the
prohibition we try to refute when refuting $O$ under $A$.~\eoe
\end{example}

\subsection{Scope and Applications}

We discuss here the scope and possible applications of
Theorem~\ref{thm:replace}.
First, Theorem~\ref{thm:replace} implies that assumptions can
undermine the separation of
obligations and prohibitions that was presented in
Lemma~\ref{lem:distinct}. This is because, under assumptions,
obligations can be replaced with prohibitions and vice versa.
But this is not surprising
when a system's set of behaviors is known to be limited (assumptions
can reflect such limitations).  For example, the obligation to turn
right at a crossroad prohibits turning left.

Second, the previous examples show that it is possible to refute
non-prohibitions under assumptions, such as totality and determinacy.
Such assumptions cannot be substantiated by testing, but may be
established using white-box techniques that analyze a system's
internal wiring or its source code.

Third, it is possible to refute
some non-prohibitions for some systems using purely
black-box analysis.  Namely, we can use Theorem~\ref{thm:replace} to
reduce the testing of an irrefutable requirement $R$ to a black-box
refutation and a black-box verification. 
For our reduction, Ted (the tester) first formulates an assumption~$A$
based on possibly unreliable information he has about the system under
test~$S$. Afterward, Ted tries to refute the hypothesis $S\in P_{R\mid
  A}$. If this succeeds, then he has shown $S\not\in R$, if $S\in
A$. In case $A$ is verifiable through tests, 
Ted tries to verify $S\in A$. If this succeeds as well,
then he has an unconditional proof of $S\not\in R$.
We illustrate this reduction  with an example.
\begin{example} 
\label{ex:combine}
In $\eio{}$, consider the requirement $R$ stating that if a system
exhibits the behavior $(1,0)$ then it should not output 0 for any
other odd input. Otherwise, it must output 0 for infinitely many
inputs. 

Clearly $R$ is not a prohibition, hence it is irrefutable. We assume
the system under test~$S$ outputs 0 for input 1, and formulate the
assumption $A$ as the set of systems that exhibit $(1,0)$.  The
prohibition $P_{R\mid A}$ then consists of the systems that do not
exhibit $0$ for any odd input larger than 1. We refute $S\in P_{R\mid
  A}$ by observing, say, $\{(3,0)\}$ in the test setup $\mathbf{T}_1$.
Again in $\mathbf{T}_1$, we verify that $S$ exhibits $(1,0)$. If these
steps succeed, we obtain $S\in A$ and $S\not \in P_{R\mid A}$. These
together entail $S\not\in R$, due to Theorem~\ref{thm:replace}.

Note that, in accordance to the results of~\S\ref{sec:semiref},
testing can refute $R$'s satisfaction only for some systems that
violate $R$.~\eoe
\end{example}

Investigating practical applications of combining black-box refutation
with black-box verification, following the above reduction, remains as
future work. In~\S\ref{sec:practice}, we present other applications of
refutability under auxiliary assumptions.

Fourth, dual to Theorem~\ref{thm:replace}, assumptions can also
facilitate verification through black-box testing. For instance, 
\emph{regularity} and \emph{uniformity} assumptions enable testers
to execute a limited number of tests and generalize their findings to
a large section of, or even the entire, input
domain~\cite{gaudel,survey}.  These assumptions enable testers to
conclude that a system satisfies a requirement by observing the
system's behaviors for a limited set of inputs.

We conclude this section with a side remark.
We have previously explored the importance of determinacy assumptions, but
obviously determinacy cannot be assumed for systems that are
known to be non-deterministic. Weaker assumptions are needed for such
systems.  For example, the \emph{complete test assumption} states that
there exists a number $n$ where if a test is executed $n$ times, then
all non-deterministic system behaviors are observed for that
particular test; see, for
instance,~\cite{CAVALCANTI20151,DBLP:journals/ipl/Hierons06}. This
assumption can be justified if there is a fairness constraint on
non-determinism, meaning that the system exercises all available
non-deterministic internal choices in a fair manner, e.g.~by
tossing a fair coin.  The complete test assumption enables testers to,
in effect, treat a non-deterministic system as a deterministic one by
repeated testing.

\section{Testing Practice Revisited}
\label{sec:practice}
Below, we review three prominent testing techniques, namely functional
testing, model-based testing, and white-box fuzz testing, and
examine them in light of our theory. This review serves two
purposes. First, it demarcates the scope and reach of our
theory. Second, it illustrates the theory's applications.
For example, we explicate the implicit assumptions 
upon which functional testing is typically based.

\paragraph{Black-box Functional Testing.} 
Functional testing refers to testing functional requirements. A
requirement is deemed {\bf functional} if it prescribes a system's
desired functionality or features.
For example, a functional requirement states that systems must have a
feature where by entering a client's identification number, a system
user obtains the client's phone number.
Functional requirements generally correspond to obligations, which are
irrefutable.
\footnote{Here, we focus our attention on those requirements that
  oblige a certain system functionality or feature. Functional
  requirements that are not obligations, e.g.\ exception-freeness,
  are exempt from our discussion. We remark that there is no canonical
  definition of functional, as opposed to non-functional, requirements
  in the literature~\cite{Glinz07}.}
Yet, functional testing is common 
practice~\cite{pezze,myers11}. This seeming discrepancy can be
resolved by noticing that the practice of functional testing is based
upon (implicit) auxiliary assumptions, as the following example
illustrates.
\begin{example}
\label{ex:phone}
Consider the $\eio{}$ model. The functional requirements $R$ obliges
systems to output $p_c$ for input $c$, where $c$ is a client
identification number, and
$p_c$ is the client $c$'s phone number.
Testing $R$ in practice amounts to providing the system under test~$S$
with the input $c$. If $S$'s output differs from $p_c$ (crash is
one such output), then the tester concludes that $S$ violates $R$.
This conclusion is justified under the assumption that $S$ is
deterministic: only then observing any behavior other than $(c,p_c)$
demonstrates $S\not\in R$.
To make $R$ refutable, the totality assumption is also needed here;
recall Example~\ref{ex:orefut}.~\eoe
\end{example}
As the example suggests, assuming that the system under test is
deterministic facilitates refuting functional requirements.
The determinacy assumption is in fact implicit in much of testing
practice.  For example, if a program passes the test that checks the
output when the input is the empty list, then most test engineers
would take this as a proof that the program behaves correctly on
empty lists. This reasoning hinges upon the assumption that the
program is deterministic, which may or may not be justified in the case
at hand.

We can now say that functional requirements are in practice refuted
under the implicit assumption of  determinacy. That is, functional
requirements are taken to forbid all outputs other than those prescribed
for a given input. For
instance, in Example~\ref{ex:phone}, for the input $c$, any output
different from $p_c$ is forbidden. However, not all obligations admit
this interpretation, as the following example illustrates.
\begin{example}
\label{ex:slot}
Consider a coffee vending machine~$V$, and a slot machine~$S$.
Suppose that~$V$ is required to output a cup of coffee and~$S$ is
required to output a winning combination, after inserting a coin. Both
these requirements are obligations. We expect a coffee machine to be
deterministic, and therefore (implicitly) add to its obligation the
following requirement:~$V$ may exhibit no behavior other than
outputting coffee. This is a prohibition and can in fact be refuted.
A slot machine, in contrast, is non-deterministic. It would be absurd
to forbid~$S$ from exhibiting non-winning combinations.  Hence, the
hypothesis that~$S$ satisfies its requirement remains
irrefutable through tests.~\eoe
\end{example}

The success of day-to-day functional testing indicates that assuming
determinacy is by and large justified in practice. Without explicating
such implicit assumptions, however, their existence and role remain
obscure. Moreover, only after explicating the determinacy assumption can
we see that functional tests that rely upon determinacy rely upon an
assumption that cannot be discharged through black-box analysis alone
(determinacy, being a prohibition, cannot be verified through
black-box tests).  Therefore, either more powerful
techniques, such as white-box analysis, should be employed for
justifying determinacy, or test results should be presented along with
the untested assumption.

We conclude this discussion by remarking that categorically assuming
that testing only applies to deterministic systems is unjustified: it
excludes concurrent systems and those that interact with an
environment containing humans, quantum sensors, and hardware prone to
stochastic failures.  Moreover, substantiating determinacy assumptions
is a formidable challenge in practice. For example, checking a
concurrent system's determinacy, even when the system's source code is
available, is hard; see,
e.g.,~\cite{DBLP:journals/cacm/BurnimS10}. This is because concurrent
systems behave differently in the presence of different
schedulers. Even if the system is amenable to white-box analysis, the
scheduler is typically either unavailable to testers or it is a
``black-box''. This substantially increases the complexity of finding
violations of requirements such as determinacy, atomicity, and
deadlock-freeness in concurrent systems.   Indeed, several existing
test methods for concurrent systems rely on instrumenting the system
or the scheduler to tame non-deterministic behaviors \cite{carver91,heisenbug,preemption}.

\paragraph{Model-based Testing.} 
In model-based testing, a system's obligations (and in some cases its
prohibitions) are represented as an
ideal model, for example by an extended state
machine~\cite{Utting:2006:PMT:1200168}. Refuting the hypothesis that
the system exhibits all the desired behaviors, specified by the model,
amounts to identifying a behavior in the model that is not exhibited
by the system.  This conforms to the lower-bound interpretation of
obligations: to find the white circle in Figure~\ref{fig:pos-neg-ex},
one can explore the oval, which represents all the obligatory
behaviors, and check if the system lacks any of them. This is the
central idea of model-based
tests~\cite{DBLP:conf/fortest/Tretmans08,Utting:2006:PMT:1200168}.

The above account of model-based testing runs into a discrepancy
similar to the one raised by black-box functional testing: the tester
is provided with an ideal model for the system under test, but has no
reasons to believe that the system does not exhibit more behaviors
than those observed during testing. The resolution again lies in
explicating auxiliary assumptions. We illustrate this point with an
example.

\begin{example}
\label{ex:mbt}
This example is based on the test method described
in~\cite[Chapter~5]{Utting:2006:PMT:1200168}.  Suppose the desired
behaviors for a system~$S$ are given as a deterministic finite-state
Mealy machine. At each state, the machine specifies the desired system
output for any input. It also specifies the system's next state, but
we ignore that part here. Suppose that the system's input domain
is~$\mathcal{I}$, and it is required to output some~$o_i$ for
input~$i$, with~$i\in \mathcal{I}$, at a certain state.
This requirement, which we call~$R$, is often implicitly interpreted
as a semi-monotone requirement~$O\wedge P$, where~$O$ obliges $S$ to
output~$o_i$ for input~$i$, and~$P$ prohibits $S$ from outputting
any~$o'\neq o_i$ for input~$i$ at that particular state.

Note that refuting~$S\in R$, under the assumption that~$S$ is
deterministic and total, is logically equivalent to refuting~$S\in P$:
if~$S$ violates~$P$, then it immediately violates~$R$. Conversely,
if~$S$ violates~$R$ because~$S\not\in O$, then~$S$ violates~$P$ as
well due to its determinacy and totality.
A tester can therefore focus on~$P$, which is indeed refutable through
tests; recall Theorem~\ref{thm:replace}.

The above reasoning is the basis of the test method prescribed
in~\cite{Utting:2006:PMT:1200168}: choose an execution of the Mealy
machine. If the system produces an input-output sequence different
from the one prescribed by the machine, then it violates its specified
requirement. Conversely, if the system violates the requirement, then
it is bound to deviate from the Mealy machine in at least one
execution.~\eoe
\end{example}

The example explicates the auxiliary assumptions that are necessary
for meaningful model-based testing in practice.  Such assumptions are
given elsewhere in the literature in different contexts, see for
example~\cite{brucker13,survey,DBLP:conf/fortest/Tretmans08,Utting:2006:PMT:1200168}.

\paragraph{White-box Fuzz Testing.}
As illustrated in previous sections, black-box tests cannot establish
the absence of behaviors.  This limitations applies to white-box fuzz
testing if the program source code is used only for generating test
inputs, as opposed to inferring the absence of behaviors. Namely,
after observing a set of behaviors, whose generation has been guided
by the source code, the tester is not justified in concluding that the
program exhibits no other behaviors. See,
e.g.,~\cite{KLEE,DBLP:journals/queue/GodefroidLM12}.
An analogous argument shows that unit testing, as in
JUnit~\cite{junit}, cannot establish the absence of behaviors if the
source code, although \emph{accessible} to testers, is not inspected
for demonstrating the absence of behaviors.

Fuzz testing is typically concerned with refuting generic
prohibitions, such as \emph{the system does not access unallocated
  memory for any input}~\cite{fuzzing-book}. To refute such
requirements, a white-box fuzzing tool covers as exhaustively as
possible the program code of the system under
test~\cite{KLEE,DBLP:journals/queue/GodefroidLM12}.  This conforms to
the upper-bound interpretation of prohibitions: to refute a
prohibition, one looks for a system behavior that is forbidden
by the prohibition.  Returning to Figure~\ref{fig:pos-neg-ex}, one
explores the triangle (which represents the set of system behaviors)
to find the black circle where the triangle intersects with
the hatched area (forbidden behaviors).

The above line of reasoning also sheds light on the suitability of the
approximation techniques that are common in (white-box) static
program analysis.  For example, a \emph{may} summary over-approximates
a program's set of behaviors~\cite{nielson}. If this does not contain
a set of obligatory behaviors, then the program violates the
corresponding obligation.  Similarly, a \emph{must} summary
under-approximates a program's set of behaviors~\cite{nielson}. If
this intersects a set of prohibited behaviors, then the program
violates the corresponding prohibition.
Note that none of these approximations is immediately applicable to
refuting requirements that are not semi-monotone because such
requirements do not admit the lower-bound and upper-bound
interpretations, as discussed in~\S\ref{sec:reqtype}.

\section{Related Work}
\label{sec:relwork}
Testing is a broad domain. We group the most closely related work into
three areas, which we present below. This complements the related work
discussed in previous sections.

\paragraph{Refutability and Verifiability.}
Our definition of refutability is inspired by Popper's notion of
\emph{testable} theories~\cite{popper}.  Theories of black-box testing
proposed in the software engineering literature are largely concerned
with the notions of test selection, test adequacy, and
exhaustiveness; see,
e.g.,~\cite{6312836,Weyuker,gaudel,DBLP:conf/fortest/Tretmans08,DBLP:journals/stvr/AiguierAGL16}.
Refutable requirements have not been investigated in prior work,
except for temporal properties and hyper-properties, which we
discussed in~\S\ref{sec:hyper} and~\S\ref{sec:refvsenf}.

Tests for \emph{verifying} the correctness of programs have been
studied in the literature; see, for
example,~\cite{howden78,budd80,Zhu97}.  The correctness guarantees
that such tests provide are inherently different from the
verifiability of obligations (\S\ref{sec:verif}), as they are reliable
guarantees only when programs and their faults satisfy assumptions
that cannot be justified solely through black-box analysis.  

Finally, tests for obtaining probabilistic correctness guarantees,
investigated for example in~\cite{focs94}, fall outside the scope of
this paper.

\paragraph{Obligations and Prohibitions.}
Obligations and prohibitions, as requirement types, implicitly appear
in various domains of software engineering.  For example, Damm and
Harel introduce \emph{existential} charts for specifying the
obligatory behaviors of a system, and \emph{universal} charts for
specifying all the behaviors the system
exhibits~\cite{DBLP:journals/fmsd/DammH01}.  An existential chart
intuitively corresponds to an obligation, and a universal chart
corresponds to a semi-monotone requirement in our theory, which is the
conjunction of an obligation and a prohibition. The notions of
necessity and possibility also have a central role in modal logic. For
example, Larsen and Thomsen's modal transition systems specify
obligations and prohibitions through, respectively, \emph{must} and
\emph{may} transitions~\cite{DBLP:conf/lics/LarsenT88}.  Similarly,
Tretmans' testing theory~\cite{DBLP:conf/fortest/Tretmans08} is based
on specifications that define both a lower-bound and an upper-bound on
a system's behaviors, which roughly speaking correspond to,
respectively, obligations and prohibitions.  These works define
prohibitions and obligations in concrete modeling formalisms. In
contrast, we present abstract definitions that can be instantiated by
the existing ones.

Finally, security requirements are sometimes called
\emph{negative}~\cite{mcgraw2006}
and \emph{universal}~\cite{tap2011}
because they do not endow a system with features and functions;
rather, they define the system's permitted behaviors. They are simply
prohibitions.

\paragraph{Testability.} 
The notion of \emph{testability} is widely used in software
engineering.  The IEEE glossary of software engineering
terminology~\cite{glossary} defines testability as: ``(1) The degree
to which a system or component facilitates the establishment of test
criteria and the performance of tests to determine whether those
criteria have been met. (2) The degree to which a requirement is
stated in terms that permit establishment of test criteria and
performance of tests to determine whether those criteria have been
met''.  Condition~(1) qualifies systems, and condition~(2)
requirements.  Our definition of refutability applies to
condition~(2). An instance of irrefutability due to failure to meet
condition~(1) is a system with unobservable error states.  Such
considerations fall outside the scope of our theory, which is built
around observations.

\section{Concluding Remarks}
\label{sec:conc}
We have formalized a simple abstract model of systems and
requirements, upon which we have built a theory of testing. Our theory
is centered around elementary notions, such as satisfiability,
refinement, and observations, and it allows us to reason precisely
about the limits and methods of black-box testing. We have used it to
fully characterize the classes of refutable and verifiable
requirements for black-box tests.  We have also clarified testing
folklore and practice.  For example, we have shown that non-exhaustive
testing can be used to verify obligations. And methodologically it
becomes clear that functional requirements can be tested only based on
assumptions that are not themselves verifiable through black-box
tests.

Our focus has been on black-box testing, defined in a general way  that
encompasses different concrete testing techniques, and its extension to
certain types of gray-box analysis. Naturally black-box tests can be
combined with other analysis techniques, like static analysis. The
indistinguishability condition of~\S\ref{sec:bbt}, stating that the
system under test can be any abstraction of an observation obtained
through tests, would then no longer be applicable.  For instance, if
the system under test is known to be deterministic, then clearly more
requirements become refutable, as discussed in~\S\ref{sec:aux}. It is
not surprising that augmenting black-box analysis with knowledge that
itself cannot be verified through black-box tests expands the
analysis's capabilities.  This paves the way for more powerful
refutation methods capable of refuting more requirements.  Developing
such an extension of our theory, and exploring its
applications remain as future work.

We remark that our theory of black-box tests is not readily
applicable to probabilistic constraints.  For example, a gambling
regulation requiring that slot machines have a 95\% payout cannot be
refuted through black-box test. Nevertheless, tests refuting such
probabilistic constraints with a controllable margin of error can be
devised. Developing a corresponding theory of tests and refutation
also remains as future work.

\bibliographystyle{plain}      
\bibliography{bib} 

\appendix
\section{Proofs}
\label{app-proof}
We present the proofs of the lemmas and theorems given in the paper.

\begin{proof}[Lemma~\ref{lem:distinct}] 
If no system satisfies~$R$, then~$R$ is trivial.  If some system~$S$
satisfies~$R$, then every system in~$\floor{\ceiling{S}}$
satisfies~$R$ because~$R$ is an obligation and a
prohibition. As~$\floor{\ceiling{S}}=\Sys$, for
any~$S\in\Sys$, we conclude that every system
satisfies~$R$. That is,~$R$ is trivial.
\end{proof}

\begin{proof}[Lemma~\ref{lem:semi}]
We split the proof of the ``iff'' claim
into two parts.

(1) Assume~$R$ is semi-monotone. We show that~$R=\ceiling{R}\wedge
\floor{R}$. Clearly~$R\subseteq \ceiling{R}\wedge \floor{R}$, for any
requirement~$R$. All we need to prove then is that~$\ceiling{R}\wedge
\floor{R}\subseteq R$. If~$\ceiling{R}\wedge \floor{R}=\emptyset$,
then the claim trivially holds. Suppose~$S\in \ceiling{R}\wedge
\floor{R}$ for some system~$S$. From~$S\in\ceiling{R}$, we
conclude~$\exists S_-\in R.\ S_-\preceq S$. Similarly,
from~$S\in\floor{R}$, we conclude~$\exists S_+\in R.\ S\preceq
S_+$. In short, we have
\begin{equation}
\label{e:1}
S_-\in R,\; S_+\in R,\; \mbox{and}\;S_-\preceq S\preceq
S_+\tag{$\dagger$} \, .
\end{equation}

Now, since $R$ is semi-monotone, one of the following three statements
holds: (a) $R$ is the conjunction of two obligations, (b) $R$ is the
conjunction of two prohibitions, or (c) $R$ is the conjunction of an
obligation $O$ and a prohibition $P$.
Case (a) along with Statement~\eqref{e:1} imply $S\in R$. The same
holds for case (b). We now consider case (c):
from $S_-\in R$ and $S_-\preceq S$ of Statement~\eqref{e:1}, we
conclude $S_-\in O$, and hence $S\in O$. Similarly, from $S_+\in R$
and $S\preceq S_+$ of Statement~\eqref{e:1}, we conclude $S_+\in P$, and
hence $S\in P$. Finally, $S\in O$, $S\in P$, and $R=O\wedge P$ imply
$S\in R$.
Therefore, if~$R$ is semi-monotone, then~$R=\ceiling{R}\wedge
\floor{R}$.

(2) Now, assume~$R=\ceiling{R}\wedge \floor{R}$. We show that $R$ is
semi-monotone. Note that for any requirement~$R$, $\ceiling{R}$ is an
obligation, hence monotone. Moreover,~$\floor{R}$ is a prohibition,
hence monotone. Therefore,~$\ceiling{R} \wedge\floor{R}$ is
semi-monotone, that is the intersection of two monotone requirements,
for any requirement~$R$. This completes the proof.

As a side note: an argument similar to (1) above shows that
$\bigwedge_{R\in \rho} R$ is semi-monotone for any nonempty set $\rho$
of semi-monotone requirements.
\end{proof}

\begin{proof}[Theorem~\ref{thm:test-neg}] 
Suppose~$R$ is $\mathbf{T}$-refutable, with
$\mathbf{T}=(T,\alpha)$. We prove that~$R$ is a prohibition.
If~$R$ is empty, then~$R$ is a trivial prohibition. If~$R$ is
nonempty, then let~$S\in R$. Now, suppose~$S'\preceq S$. All we need
to prove is that~$S'\in R$, which we prove by contradiction.

Assume~$S'\not\in R$. Then~$\exists t\in T.\ \hat\alpha(t)\cap
R=\emptyset$ simply because~$R$ is
$\mathbf{T}$-refutable. Since~$\alpha$ is order-preserving
and~$S'\preceq S$, we have~$t\in \alpha(S)$. Therefore,~$S\in
\hat\alpha(t)$. This entails~$S\not\in R$, which contradicts the
assumption~$S\in R$. We conclude that~$S'\in R$. Therefore,~$R$ is a
prohibition.
\end{proof}

\begin{proof}[Lemma~\ref{lem:compos}]
Let $\mathbf{T}=(T,\alpha)$, and write $W$ for
$\bigwedge_{R\in\rho}R$. Suppose a system $S$ violates $W$. Then
there is at least one $R\in \rho$ such that $S\not\in R$. Since $R$ is
$\mathbf{T}$-refutable, there is an observation $t\in \alpha(S)$ such
that $\hat\alpha(t)\cap R=\emptyset$. Now, from $W\subseteq R$ we
obtain $\hat\alpha(t)\cap W=\emptyset$. This shows that $W$ is
$\mathbf{T}$-refutable.
\end{proof}

\begin{proof}[Lemma~\ref{lem:most-permissive}]
Fix a system model~$\model=(\Sys,\preceq,\bot,\top)$, and
let~$R$ be a prohibition. We show that~$R$
is~$\mathbf{T}_r^\model$-refutable, where
$\mathbf{T}_r^\model=(\Sys,\floor{\cdot})$.

Assume that some system~$S$ violates~$R$. Since~$R$ is a prohibition,
any system that abstracts~$S$
violates~$R$. Moreover,~$S\in\floor{S}$. We conclude that~$\exists
S_w\in\floor{S}.\ \ceiling{S_w}\cap R=\emptyset$, namely $S_w=S$.
Hence $R$ is~$\mathbf{T}_r^\model$-refutable.
\end{proof}

\begin{proof}[Lemma~\ref{thm:pos-not-ref}]
Suppose~$R$ is a nontrivial obligation. We prove by contradiction that
$R$ is not refutable in any test setup.

Assume that~$R$ is $\mathbf{T}$-refutable in some test
setup~$\mathbf{T}$. By Theorem~\ref{thm:test-neg},~$R$ is a
prohibition. Then,~$R$ must be trivial by Lemma~\ref{lem:distinct},
because~$R$ is both a prohibition and an obligation. That~$R$ is
trivial contradicts the assumption that $R$ is a nontrivial
obligation. Hence $R$ is not refutable in any test setup.
\end{proof}

\begin{proof}[Theorem~\ref{thm:dual}]
Suppose~$R$ is $\mathbf{T}$-verifiable, with
$\mathbf{T}=(T,\alpha)$. We prove that~$R$ is an obligation.  If~$R$
is empty, then~$R$ is a trivial obligation. If~$R$ is nonempty, then
let~$S\in R$.  Now, suppose~$S\preceq S'$.  All we need to prove is
that~$S'\in R$.  Since~$R$ is $\mathbf{T}$-verifiable, from~$S\in R$
we conclude~$\exists t\in\alpha(S).\ \hat\alpha(S)\subseteq
R$. As~$\alpha$ is order-preserving and~$S\preceq S'$, we have~$t\in
\alpha(S')$. That is,~$S'\in\hat\alpha(S)$. We conclude that~$S'\in
R$. Therefore,~$R$ is an obligation.
\end{proof}

\begin{proof}[Lemma~\ref{lem:prohver}]
Suppose~$R$ is a nontrivial prohibition.  We prove by contradiction
that $R$ is not verifiable in any test setup.

Assume that~$R$ is $\mathbf{T}$-verifiable in some test
setup~$\mathbf{T}$.  By Theorem~\ref{thm:dual},~$R$ is an obligation.
Then,~$R$ must be trivial by Lemma~\ref{lem:distinct}, because~$R$ is
both a prohibition and an obligation.  That~$R$ is trivial contradicts
the assumption that $R$ is a nontrivial prohibition. Hence
$R$ is not verifiable in any test setup.
\end{proof}

\begin{proof}[Lemma~\ref{lem:temporal}]
We split the proof into three parts, reflecting the lemma's 
claims.

(1) Let~$\phi$ be a~$\mathbf{T}_*$-refutable property. We show
that~$\phi$ is safety.

Assume~$\pi\not\in\phi$, for some~$\pi\in\Sigma^\omega$. Then,
the system~$S_\pi=\{\pi\}$ violates~$\phi$.
Now, by~$\phi$'s~$\mathbf{T}_*$-refutability, there exists a finite
set~$t$ of~$\phi$'s finite prefixes that demonstrates~$S_\pi\not\in
R_\phi$, where $R_\phi=\floor{\phi}$. Let~$\sigma$ be the longest
element in~$t$; note that since~$\{\pi\}$ is a singleton, there always 
exists a single longest element in~$t$. Then, for any~$\pi'\in
\Sigma^\omega$, the system~$S_{\pi'}=\{\sigma\pi'\}$ violates~$\phi$,
simply because~$t$ belongs to~$\alpha(S_{\pi'})$. We conclude
that~$\sigma\pi'\not\in\phi$, for all~$\pi'\in\Sigma^\omega$. That
is,~$\phi$ is a safety temporal property.

(2) Let~$\phi$ be a safety property. We show that~$\phi$
is~$\mathbf{T}_*$-refutable.

Assume that a system~$S$ violates~$\phi$. That is,~$\exists \pi\in
S.\ \pi\not\in\phi$. Since~$\phi$ is safety, a finite prefix of~$\pi$,
say~$\sigma$, satisfies the following condition:~$\forall
\pi'\in\Sigma^\omega.\ \sigma\pi'\not\in\phi$. Now, define the
observation~$t\in T_*$ as~$\{\sigma\}$. Note that~$t\in\alpha(S)$, and
moreover~$\hat\alpha(t)\cap R_\phi=\emptyset$ due to the above
condition. This shows that~$\phi$ is~$\mathbf{T}_*$-refutable.

(3) Any temporal property~$\phi$ is $\mathbf{T}_r$-refutable
  because~$R_\phi$'s satisfaction is refinement-closed for
  any~$\phi$. Then, by Lemmas~\ref{lem:distinct}
  and~\ref{lem:prohver}, any $\mathbf{T}_r$-verifiable or
  $\mathbf{T}_*$-verifiable property must be trivial.
This completes the proof.
\end{proof}

\begin{proof}[Theorem~\ref{thm:replace}]
Let $P=\floor{A\cap R}$.  That $P$ is a prohibition is
immediate. Below, we prove the contrapositive form of the
statement $\forall S\in A.\ S\not\in P \leftrightarrow S\not\in R$ in
two directions.

(1) We show $\forall S\in A.\ S\in R\to S\in P$.  Let $S$ be a system
in $A$ that satisfies $R$.  Then, $S\in A\cap R$, and hence $S\in P$.

(2) We show $\forall S\in A.\ S\in P\to S\in R$.  Let $S$ be a system
in $A$ that satisfies $P$. We assume $S\not\in R$, and derive a
contradiction as follows.
From $S\in P$, we conclude that there is a system $S'$ such that
$S\preceq S'$ and $S'\in A\cap R$.
Since $R$ is $\mathbf{T}$-refutable under $A$, and $S\in A$, there is
an observation $t\in\alpha(S)$ such that $\hat\alpha(t)\cap A\cap
R=\emptyset$. As $S\preceq S'$, we have $S'\in \hat\alpha(t)$.

From the above results we conclude $S'\in \hat\alpha(t)\cap A \cap R$,
which contradicts $\hat\alpha(t)\cap A\cap R=\emptyset$. Therefore,
$S\in R$, which completes the proof.
\end{proof}

\end{document}